\documentclass[preprint,superscriptaddress]{revtex4-1}

\usepackage[british,UKenglish,USenglish,english,american]{babel}
\DeclareMathAlphabet{\mathpzc}{OT1}{pzc}{m}{it}
\usepackage{graphicx}
\usepackage{capt-of}
\usepackage{caption}
\usepackage{epsfig,color}
\usepackage[T1]{fontenc}
\usepackage{mathptmx,avant,courier}
\usepackage{amsmath}
\usepackage{calligra}
\usepackage[version=3]{mhchem}
\usepackage{mathrsfs}
\usepackage{paralist}
\usepackage[symbol]{footmisc}

\usepackage[a4paper, left=1cm, right=1cm, top=1cm, bottom=1cm]{geometry}
\begin{document}

\title{Skyrme-Hartree-Fock calculations of nuclear properties in the drip-point region of neutron star crust}

\author{Uwe Heinzmann}
\affiliation{Frankfurt Institute for Advanced Studies, Goethe University, D-60438 Frankfurt am Main, Germany}
\author{Igor N. Mishustin}
\affiliation{Frankfurt Institute for Advanced Studies, Goethe University, D-60438 Frankfurt am Main, Germany}
\affiliation{National Research Center Kurchatov Institute, Moscow 123182, Russia}
\author{Stefan Schramm\dag \footnote[0]{$\dag$  deceased}}
\affiliation{Frankfurt Institute for Advanced Studies, Goethe University, D-60438 Frankfurt am Main, Germany}
\affiliation{Institute for Theoretical Physics, Goethe University, D-60438 Frankfurt am Main, Germany}
\date{\today} 

\begin{abstract}
In the present paper we explore the neutron-drip region of cold non-rotating isolated neutron stars. We have performed extended nuclear-structure calculations for nuclei embedded in the electron gas. For modeling the outer crust we use a set of Wigner-Seitz cells, where every cell contains one nucleus surrounded by a cloud of relativistic electrons. Above the drip point a non-relativistic neutron gas occurs in the cell. These calculations are carried out within the Hartree-Fock approach in combination with Skyrme effective interactions. For every baryon density we have determined the configuration with a minimal total energy. The drip elements and corresponding drip densities have been determined for about 240 different parametrizations of Skyrme forces used in the literature. We demonstrate that the calculated drip-point densities depend essentially on the Skyrme parametrization used. Even the drip elements and the occupied shells in the drip region differ for different parametrizations. We have found that the number of neutrons building the neutron gas at the drip point also depends essentially on the Skyrme force chosen. Nevertheless, the number density of the neutron gas in the drip-region is more or less the same ($\sim 10^{-5} \frac{neutrons}{{fm}^{3}}$). The drip densities obtained within our approach are generally lower than predicted earlier. 
\end{abstract}
\maketitle
\thispagestyle{empty}

\newpage
\section{Introduction}
After detecting gravitational waves from neutron-star mergers, compact objects are now again in the focus of interest. From these catastrophic events one can obtain important information about the properties of strongly-interacting matter in a very broad domain of baryon densities and temperatures. In the present paper we focus on cold isolated non-rotating neutron stars (NS), which have mass typically of the order of a solar mass ($M_\odot=1,29\cdot 10^{30}$kg) and radii of about 10 kilometers, see e.g. review \cite{Camenzind07}. They have only a small atmosphere and a thin solid crust forming the outer layers. The thickness of the crust is about 1 km and the mass contained in the crust is about one percent of the total mass. But nevertheless, its structure is relevant for the interpretation of many observational data. For example glitches, sudden changes of the pulsar periods, are thought to be caused by the breakdown of the crust due to the slowing down of the rotation. Even more violent destruction of the crust is expected in NS-NS merger events, see \cite{Tsang}. Also the cooling rate of magnetars is rather sensitive to the composition and the thickness of the crust, see e.g. \cite{AquMi08}. Generally, the whole information coming from inner layers of the neutron star is filtered by the crust material, see more in refs. \cite{Becker09}, \cite{Glend00}, \cite{Glend07}.\\
One usually divides the crust into two regions: the outer crust and the inner crust. The outer part of the crust has a crystalline structure consisting of more or less spherical nuclei surrounded by electrons. The inner crust is defined as the transition layer between the neutron drip density $\rho_{drip}$ and the homogeneous nucleonic matter (outer core) at about half of the nuclear saturation density, $\rho_{0} \approx 0.16 \mbox{ fm}^{-3}$. At density $\rho_{drip}$ the neutron chemical potential becomes equal to $m_{n}c^{2}$ and neutrons drip out from the nuclei. At higher densities the nuclei are surrounded by a neutron gas whose pressure together with electrons acts against gravity.\\
In this paper we use fully microscopic Hartree-Fock approach to determine corresponding drip nuclei and their neighbours. These calculations are done for a large variety of Skyrme effective interactions used in the literature. More specifically, we have considered about 240 different Skyrme parametrizations, which have been selected and analyzed in ref. \cite{DuSto12}. Most previous calculations were done for one specific Skyrme force without analysing the differences. According to our results, the neutron drip point is more likely between $3,0\cdot 10^{11} \frac {\mbox{g}}{\mbox{cm}^{3}}$ and $4,0\cdot 10^{11} \frac{\mbox{g}}{\mbox{cm}^{3}}$ than between $4,0\cdot 10^{11} \frac {\mbox{g}}{\mbox{cm}^{3}}$ and $5,0\cdot 10^{11} \frac{\mbox{g}}{\mbox{cm}^{3}}$, as found previously. Assuming a neutron drip at a lower density would cause a thinner outer crust and, consequently, a thicker inner crust of the neutron star. Obviously, this  may lead to significant phenomenological consequences, such as the tidal deformability of neutron stars in merging events, see  e.g. refs. \cite{Biswas}, \cite{Gittins}.   

\section{Previous estimates of drip-point density}

\subsection{Early estimates}
First estimates of the drip density and drip elements have been done in ref. \cite{BPS71}. The authors described the outer crust within a thermodynamic approach using a phenomenological equation of state obtained by extrapolating of known nuclear mass data. By minimizing the total energy of the system of the nucleus, the electrons and their interaction (lattice energy), they estimated the drip density to be $4,3 10^{11} \frac{\mbox{g}}{\mbox{cm}^{3}}$ with the drip nucleus $^{118}Kr$.\\
Later Negele and Vautherin \cite{NegVauth73} described the properties of the inner crust of a neutron star using energy-density functional method combined with the Wigner-Seitz cells \cite{WS33,WS34}. Within this approach the crystal structure of the crust is approximated with a set of independent spherical cells, each containing one nucleus (N,Z) and Z electrons. Electrons form a degenerate Fermi gas, which becomes fully relativistic at a density of $\rho \sim 10^{7} \frac{\mbox{g}}{\mbox{cm}^{3}}$. Hence, at densities around the neutron drip point the crystal consists of more or less spherical nuclei immersed in an uniform gas of ultra-relativistic electrons.

\subsection{Estimates using Liquid-drop model}

Below we ignore small corrections due to deviations from spherical symmetry of electron distributions. The volume per nucleus $n_{N}$ is defined using a sphere of radius $r_{C}$:
\begin{equation}
  n_{N}= \frac{1}{\frac{4}{3} \pi r_{C}^{3}},
\end{equation}
where $r_{C}$ is in the order of magnitude of the nuclear spacing. The electron density $n_{e}$ is defined in a similar way as
\begin{equation}
  n_{e}= \frac{1}{\frac{4}{3} \pi r_{e}^{3}},
\end{equation}
defining a corresponding length $r_{e}$. Because of electrical neutrality the equation
\begin{equation}
  n_{e}= Z \cdot n_{N}
\end{equation}
must hold. Therefore the two lengths are correlated as
\begin{equation}
  r_{c}=Z^{\frac{1}{3}}\cdot r_{e}.
\end{equation}
To get a rough estimate of the neutron drip density, we use a simple liquid drop model, as was first done in ref. \cite{PethRav91}. The total energy of a nucleus with mass number $A$ and proton number $Z= x\cdot A$ can be expressed as
\begin{equation}
E^{\mbox{nucl}}_{\mbox{tot}}(A,Z) = (A-Z)\cdot m_{n} + Z\cdot m_{p} - a_{V} \cdot A + 
      + a_{sym}\cdot (1-2x)^{2}\cdot A + a_{S} \cdot A^{2/3} + a_{C}\cdot x^{2} \cdot A^{5/3}, 
\end{equation}
where $A=Z+N$, $m_{n}$ and $m_{p}$ are the rest masses of neutron and proton, $a_{V}$, $a_{sym}$, $a_{S}$ and $a_{C}$ are the corresponding empirical coefficients. We have used units $\hbar=c=1$.\\

The total energy per nucleon of the whole system including a nucleus and $Z$ electrons can be written as
\begin{equation}
 \frac{E_{\mbox{tot}}}{A} = x\cdot \epsilon_{e}+ \frac{E^{\mbox{nucl}}_{\mbox{tot}}(A,Z)}{A}, 
\end{equation}
  where $\epsilon_{e}$ is the average electron energy per particle.
  To find the optimum nucleus (A*,Z*) one should  minimize this expression, first with respect to $A$, and then, with respect to $x$. The first minimization yields
\begin{equation}
a_{s}A^{2/3} = 2 \cdot a_{C} x^{2}A^{5/3} \mbox{       or          } A^*=\frac{a_{s}}{2a_{C}x^{2}},
\end{equation}
meaning that the optimal nucleus has surface energy per nucleon equal {\bf twice} its Coulomb energy. Here one can see that with increasing nuclear mass, i.e. $A$, the proton fraction $x=\frac{Z}{A}$ of the optimal nucleus decreases. The next step is to find the optimal proton fraction. After substituting the expression for $A^*$ into $E_{\mbox{tot}}/A$, one can differentiate it with respect to $x$ at fixed $N$. Rewriting the sum of surface and Coulomb energies per nucleon as $(\frac{3}{2})\cdot \left( 2a_{s}^{2}a_{C}x^{2}\right)^{1/3}$, one finally gets 
\begin{equation}
 \mu_{e} + m_{p} - m_{n} - 4a_{sym}\cdot (1 - 2x) + \left(2a_{C}a_{s}^{2}\cdot x^{2}\right)^{1/3}=0,
\end{equation}
were $\mu_{e}=\frac{\partial E_{e}}{\partial N_{e}}$ is the electron chemical potential. This equation is another representation of the {\bf thermodynamic equilibrium condition} $\mu_{e} + \mu_{p} = \mu_{n}$, where $\mu_{p}$ and $\mu_{n}$ are the proton and neutron chemical potentials including the corresponding rest masses. This simply means that under the thermodynamic equilibrium, electron capture and neutron decay are in chemical equilibrium. Now this equation relates the electron chemical potential to the proton fraction. At densities about $10^{11}\frac{\mbox{g}}{\mbox{cm}^{3}}$, where the neutron drip starts, the electrons can be treated as an ultra-relativistic Fermi gas, e.g. $\mu_{e}=p_{Fe}$, where $p_{Fe}$ is the electron Fermi momentum, related to the electron density as $n_{e} = \frac{1}{3\pi^{2}}(\mu_{e})^{3}= x \cdot n$, where $n=\frac{A}{V}$ is the nucleon density. Substituting this in (8) gives $x$ as a function of $n$ or alternatively of the mass density $\rho=m\cdot n$, where $m = \frac{1}{2}(m_{n} + m_{p})$ is the average nucleon mass. The physical condition for dripping out of neutrons is that the neutron chemical potential equals the neutron's rest mass, i.e. {\bf $\mu_{n} = m_{n}$}. The corresponding chemical potential is obtained by differentiating the  nuclear energy density with respect to the total neutron number density, $n_{n}=(1-x)\cdot n$. The result of this differentiation is 
\begin{equation}
\mu_{n}=\left(\frac{\partial E^{\mbox{nucl}}_{\mbox{tot}}(A,Z)}{\partial N}\right)_{Z} = m_{n} - a_{V} + a_{sym} \cdot (1-4x^{2}) + \frac{1}{2}(2a_{C}a_{s})^{1/3}\cdot x^{2/3}.
\end{equation}
Therefore, the condition for neutron drip reads
\begin{equation}
a_{V} = a_{sym}\cdot (1 - 4 x^{2}_{drip}) + \frac{1}{2}(2\cdot a_{C}a^{2}_{s})^{1/3} x_{drip}^{2/3}.
\end{equation}
Neglecting the surface term and Coulomb corrections (second term) yields $x^{2}_{drip}=\simeq \frac{1}{4}\left(1 - \frac{a_{v}}{a_{sym}}\right)$. Using the values $a_{V} = 15.5$ MeV and $a_{sym} = 23$ MeV from ref. \cite{Bethge96}, one gets $x_{drip} = 0.2855$. Inserting these values in (7) yields $A_{drip} \simeq 144$ and thus $Z_{drip} \simeq 41$, which is the {\bf Niobium isotope $_{41}^{144}\mbox{Nb}$}. Pethick and Ravenhall \cite{PethRav91} used somewhat different parameters $a_{V} \simeq 16$ MeV and $ a_{sym} \simeq 24$ MeV and obtained the value $x_{drip} = 0.32$. Therefore, they got $A_{drip} \simeq 122$ and $Z_{drip} \simeq 39$ which is the {\bf Yttrium isotope $_{39}^{122}\mbox{Y}$}.\\

By using $\mu_{e}$ from the exact equation (8), after inserting $a_{sym} =23$ MeV, $a_{C}=0.715$ MeV and $a_{s} = 16.8$ MeV, one gets $\mu_{e}=29.544\mbox{ MeV}$, which is $\sim 5$ MeV higher than the value quoted by Pethick and Ravenhall in ref. \cite{PethRav91}, $\mu_{e} =25$ MeV. Using the parameters of the liquid-drop model from ref. \cite{Bethge96} $a_{V}= 16$ MeV and $a_{sym}=24$ MeV one gets $n_{e} = 1.133\cdot 10^{-4} \frac{1}{\mbox{fm}^{3}}$ and $r_{e}=12.81$ fm, that is about 20\% lower than the value quoted by Pethick and Ravenhall, $r_{e}=15.14$ fm.\\

The corresponding mass density is calculated as
\begin{equation}
\rho = (A\cdot m)\cdot \frac{n_{e}}{Z}= m\cdot\frac{A}{Z} \cdot \frac{1}{\frac{4}{3}\pi r^{3}_{e}}.
\end{equation}
Finally we get for the mass density at neutron drip point $\rho_{drip}= 0.3735 \frac{\mbox{MeV}}{\mbox{fm}^{3}}$ or  $6,658 \cdot 10^{11} \frac{\mbox{g}}{\mbox{cm}^{3}}$, which is significantly higher than the result reported in \cite{PethRav91}, $3,5 \cdot 10^{11}\frac{\mbox{g}}{\mbox{cm}^{3}}$. Comparing these results one can conclude that even the parameters of the liquid-drop model do have astrophysical relevance. The situation is even more striking for calculations using multiple versions of Skyrme effective interactions, as will be demonstrated in Sect. III.
 
\subsection{Symmetry energy based estimates}

A first rough estimation of the neutron drip density using the symmetry energy $S_{0}$ has been done in ref. \cite{HaePoYa07}. The authors use a simplified version of the nuclear mass formula neglecting Coulomb, surface and all other finite-size terms. Keeping only quadratic term in $\delta = (N-Z)/A$ one can write the energy per nucleon in nucleus (A,Z) as
\begin{equation}
\frac{E_{N}(A,Z)}{A} \simeq \epsilon_{0} +  S_{0} \cdot \delta^{2},
\end{equation}
where $\epsilon_{0}$ is the bulk energy per nucleon with respect to the rest mass and $S_{0}$ is the symmetry energy coefficient for symmetric nuclear matter, both calculated at the saturation density. The neutron-proton mass difference is neglected. Following this approximation, the neutron chemical potential with respect to the neutron mass is $\mu'_{n} = \epsilon_{0} + (2\delta + \delta^{2})S_{0}$ and the corresponding proton chemical potential is $\mu'_{p} = \epsilon_{0} + (-2\delta + \delta^{2})\cdot S_{0}$. Now the value of $\delta_{D}$ corresponding to the neutron-drip density $\rho_{D}$ can be calculated from the condition $\mu'_{n}=0$ or $\delta_{D}= \sqrt{1 - \frac{\epsilon_{0}}{ S_{0}}}-1$. For example, using  values for the Skyrme force SKM* from ref. \cite{DuSto12} $\epsilon_{0} = -15.77$ MeV and $S_{0}= 30.03$ MeV, we get $\delta_{D} = 0.2349$. From the $\beta$-equilibrium condition we have $\mu_{e} = \mu_{n}-\mu_{p} \simeq 4\cdot S_{0} \cdot \delta$. On the other hand, for relativistic electrons $\mu_{e}\approx p_{Fe}=\left(3\pi^{2}n_{b}\cdot \frac{Z}{A}\right)^{1/3}$. Using this identity and converting the number density into mass density $\rho$ one gets for $\mu_{e}$:
\begin{equation}
\mu_{e}= 0.516 \cdot (\rho_{6}\cdot \frac{Z}{A})^{\frac{1}{3}},
\end{equation}
where $\rho_{6} \equiv \frac{\rho}{10^{6}}$, see details in ref. \cite{HaePoYa07}. Computing $x_{D}$ by inserting $\delta_{D}$ in $x_{D} = \frac{Z}{A}=\frac{1}{2}(1-\delta_{D})= 0.38255$ one ends up with the formula
\begin{equation}
\rho_{6_D} = \left(\frac{4\cdot S_{0}\cdot \delta_{D}}{0.516}\right)^{3} \cdot \frac{1}{x_{D}}.
\end{equation}
Putting in the values for SKM*, one obtains $\rho_{drip} \simeq 4,274\cdot 10^{11}\frac{\mbox{g}}{\mbox{cm}^{3}}$, which is surprisingly close to the drip density calculated in ref. \cite{BPS71} (BPS), $4,3 \cdot 10^{11} \frac{\mbox{g}}{\mbox{cm}^{3}}$, being now a commonly accepted value. With the nuclear saturation density $n_{0} = 0.16 \mbox{ fm}^{-3}$ equivalent to $\rho_{0}=2.4 \cdot 10^{14}\frac{\mbox{g}}{\mbox{cm}^{3}}$ even BBP ( see ref. \cite{BBP71}) obtained in their calculations for the drip density $\rho_{drip}=4,3\cdot 10^{11}\frac{\mbox{g}}{\mbox{cm}^{3}}$.\\
\\
More recently a systematic analysis of the outer crust structure in the vicinity of the drip point has been carried out in ref. \cite{RueHeSB06}. A set of different nuclear models for the nuclear equation of state has been used within a thermodynamic approach described in ref. \cite{BBP71}. The predicted densities for the neutron drip point were found around the value $\rho_{drip} = 4 \cdot 10^{11}\frac{\mbox{g}}{\mbox{cm}^{3}}$. In the drip region the authors have found a number of elements with proton numbers from {\bf Z=34} (Selenium) to {\bf Z=38} (Strontium).

\section{Realistic Skyrme-Hartree-Fock approach}

\subsection{General remarks}

The neutron drip area of neutron star crust is in the focus of our present research. For the description of nuclei we use Hartree-Fock approach in combination with Skyrme effective interactions. We consider spherical Wigner-Seitz cells containing a nucleus immersed in relativistic electron gas, and perform thousands of Skyrme-Hartree-Fock calculations running through all possible combinations of protons and neutrons. After that we determine the ground state energy of the cell including the energy of electron gas as a function of $\rho$. The most stable nuclei are determined in a broad density range up to the neutron drip point. We have found significant differences in predicted drip elements for different Skyrme parametrizations. These uncertainties should be taken into account when calculating crust properties and cooling rates.

\subsection{Skyrme energy-density functionals}

Our goal is to model the neutron star crust within the framework of a Skyrme-Hartree-Fock (SHF) approach with BCS-pairing. We assume  that only ground-state nuclei are present in the crust of neutron stars, and consider below only {\it even-even} nuclei. There are different possibilities to build a Skyrme energy density functional which are discussed in ref. \cite{RS04}. In our calculations we adopt a parametrization previously used in refs. \cite{Greiner95}, \cite{Bender97}. A general energy-density functional for an interacting finite system of neutrons (n) and protons (p) can be represented as
\begin{equation}
  {\cal E}_{{tot}}= {\cal E}_{kin}+{\mathcal E}_{Skyrme}+{\cal E}_{Coulomb}+{\cal E}_{pair}-{\cal E}_{corr}.
\end{equation}
Here ${\cal E}_{kin}$ is the kinetic energy of nucleons, calculated as ${\cal E}_{kin}=\int d^{3}r \tau(r)$, where $\tau =\tau_{p} + \tau_{n}$ is the sum of the kinetic energy densities of protons and neutrons. The particle densities are defined as
\begin{equation}
  \rho_{q}=\sum_{k\in q}\upsilon^{2}_{k}|\psi_{k}({\bf r})|^{2},
\end{equation}
where we have introduced index $q$, which denotes proton (p) and neutrons (n). Here $\upsilon_{k}$ are the variational parameters in the wave function for even-even nuclei represented in the BCS mode as $q$, see details in ref. \cite{RS04}:
\begin{equation}
|\mbox{BCS}\rangle = \prod_{k>0}(u_{k}+\upsilon_{k}\hat{a}^{\dagger}_{k}\hat{a}^{\dagger}_{\bar{k}}|\mbox{0}\rangle,  
\end{equation}
where $u_{k}^{2}+\upsilon_{k}^{2}=1$. This equation follows from the normalization condition of the BCS state
\begin{equation}
\langle\mbox{BCS}|\mbox{BCS}\rangle = N.
\end{equation}
The A-particle wave function $|\psi \rangle$ can be chosen as a superposition of all A-particle Slater determinants which can be written as
\begin{equation}
  |\psi \rangle = \sum_{i_{1},i_{2}, \dots i_{A}}c_{i_{1}}, \dots , c_{i_{A}}\cdot \hat{a}_{i_{1}}^{\dagger}\cdots \hat{a}_{i_{A}}^{\dagger}|0\rangle,
\end{equation}
where the sets $\{i_{1}, i_{2}, \dots i_{A}\}$ represent the subspace of a complete basis of one-particle states.\\

The kinetic energy densities  can be calculated as
\begin{equation}
\tau_{q}({\bf r}) = \nabla \cdot \nabla'\rho_{q}({\bf r},{\bf r}')\big|_{{\bf r} ={\bf r'}}=\sum_{k \in q}\upsilon^{2}_{k}|\nabla \psi_{k}({\bf r})|^{2}.
\end{equation}

The contribution of the nuclear mean field is represented by the general Skyrme functional
\begin{align}
\begin{split}
{\cal E}_{Skyrme}=\int d^{3}r \bigg\{\frac{b_{0}}{2}\rho^{2}-\frac{b^{'}_{0}}{2}\sum_{q}\rho^{2}_{q}+b_{1}(\rho \tau -{\bf j}^{2})-b^{'}_{1}\sum_{q}(\rho_{q}\tau_{q}-{\bf j}^{2}_{q}) \\
-\frac{b_{2}}{2}\rho\Delta\rho +\frac{b^{'}_{2}}{2}\sum_{q}\rho_{q}\Delta\rho_{q}+\frac{b_{3}}{3}\rho^{\alpha+2}-\frac{b^{'}_{3}}{3}\rho^{\alpha}\sum_{q}\rho^{2}_{q} \bigg\}+{\cal E}_{LS},
\end{split}
\end{align}
where $ {\bf j}_{q}$ are the current densities defined as
\begin{eqnarray}
{\bf j}_{q}({\bf r})&=& -\frac{\mbox{i}}{2}\left(\nabla - \nabla' \right)\rho_{q}({\bf r},{\bf r'})\big|_{{\bf r}= {\bf r'}} \nonumber \\
 &=& -\frac{\mbox{i}}{2}\sum_{k \in q}{\mathpzc v}_{k}^{2} \{\psi_{k}^{\dagger} ({\bf r})\nabla \psi_{k}({\bf r})-[\nabla \psi_{k}^{\dagger}({\bf r})]\psi_{k}({\bf r})\},
\end{eqnarray}
and the local single particle density is defined as
\begin{equation}
  \rho_{q}({\bf r})=\rho_{q}({\bf r},{\bf r})=\sum_{k \in q}{\upsilon}_{k}^{2}|\psi_{k}({\bf r})|^{2}.
\end{equation}
${\cal E}_{LS}$ is the spin-orbit interaction term (see below). The nuclear Coulomb energy is calculated in the  local density approximation including the exchange term,
\begin{equation}
{\cal E}_{Coulomb}=e^{2}\frac{1}{2}\iint d^{3}\mbox{r  }d^{3}\mbox{r}^{'}\frac{\rho_{p}({\bf r})\rho_{p}(\bf r^{'})}{|{\bf r}-{\bf r^{'}}|}-\frac{3}{4}e^{2}\left(\frac{3}{\pi}\right)^{1/3}\int d^{3}\mbox{r  }[\rho_{p}({\bf r})]^{4/3},
\end{equation}
where the {\it naked} proton density is used,  for details see ref. \cite{Greiner89b}.

Following ref. \cite{Bender97}, for the spin-orbit interaction we consider three possibilities: 
\begin{subequations}
\begin{alignat}{3}
&{\cal E}^{std}_{LS} = \int d^{3}r \bigg\{-b_{4}[\rho {\bf \nabla \cdot J}+{\bf s \cdot \nabla}\times {\bf j}+\sum_{q}(\rho_{q}{\bf \nabla \cdot J}_{q}+{\bf s}_{q}{\bf \cdot \nabla}\times {\bf j}_{q})]\bigg\},\\
&{\cal E}^{({\bf J^{2}})}_{LS}= {\cal E}^{std}_{LS}-\int d^{3}r\bigg\{ \frac{1}{16}(t_{1}x_{1}+t_{2}x_{2})\left({\bf J^{2}}-2{\bf s \cdot \tau} \right)+\frac{1}{16}(t_{1}-t_{2})\sum_{q}({\bf J^{2}}_{q}-2{\bf s}_{q}\cdot{\bf \tau}_{q})\bigg\},\\
&{\cal E}^{ext}_{LS} = \int d^{3}r \bigg\{-b_{4}(\rho {\bf \nabla\cdot J}+ {\bf s \cdot \nabla}\times{\bf j})-b^{'}_{4}\sum_{q}(\rho_{q}{\bf \nabla\cdot J}_{q}+ {\bf s}_{q}{\bf\cdot \nabla}\times {\bf j}_{q}) \bigg\}.
\end{alignat}
\end{subequations}

 Finally  ${\cal E}_{pair}$ is the pairing energy density and the term ${\cal E}_{corr}$ contains the center-of-mass correction. Higher order corrections to the exchange term do not play a role, as was shown in ref. \cite{Tit74}.

The variation of ${\cal E}_{Skyrme}$ with respect to the density $\rho_{q}$ yields the self-consistent nuclear potential   
\begin{align}
\begin{split}
U_{q}({\bf r})=\frac{\delta {\cal E}}{\delta \rho_{q}({\bf r})}=& b_{0}\rho({\bf r})-b^{'}_{0}\rho_{q}({\bf r})+b_{1}\tau({\bf r})-b_{1}^{'}\tau_{q}({\bf r})-b_{2}\Delta \rho({\bf r})+b^{'}_{2}\Delta \rho_{q}({\bf r})\\
&+b_{3}  \frac{\alpha+2}{3}\rho^{\alpha+1}({\bf r})-b^{'}_{3}\frac{2}{3}\rho_{q}^{\alpha}({\bf r})-b^{'}_{3}\frac{\alpha}{3}\rho^{\alpha-1}({\bf r})\sum_{q^{'}}\rho^{2}_{q^{'}}({\bf r})\\
&-b_{4}\nabla \cdot {\bf J}({\bf r})-b^{'}_{4}\nabla \cdot {\bf J}_{q}({\bf r}).
\end{split}
\end{align}

For the protons one adds also the Coulomb term 
\begin{equation}
U_{c}=e^{2}\int d^{3}r^{'}\frac{\rho_{p}({\bf r}^{'})}{|{\bf r}-{\bf r}^{'}|}-e^{2}\left(\frac{3}{\pi}\right)^{1/3}\rho^{1/3}_{p}({\bf r}).
\end{equation}

The corresponding spin-orbit potential is calculated as

\begin{equation}
{\bf W}_{q}({\bf r})=\frac{\delta {\cal E}}{\delta{J}_{q}({\bf r}^{'})}-\nabla \frac{\delta {\cal E}}{\delta (\nabla \cdot {\bf J}_{q}({\bf r}))}.
\end{equation}
Depending on the choice made in eq.(20), one gets three possibilities:
\begin{subequations}
\begin{alignat}{3}
&{\bf W}^{std}_{LS} = b_{4}\cdot\left(\nabla \rho ({\bf r}) + \nabla \rho_{q}({\bf r})\right),\\
&{\bf W}^{({\bf J^{2}})}_{LS}= \frac{1}{8}\cdot(t_{1}-t_{2}){\bf J}_{q}-\frac{1}{8}\cdot(t_{1}x_{1}+t_{2}x_{2})\cdot{\bf J}+b_{4}\cdot \left(\nabla \rho({\bf r})+\nabla \rho_{q}({\bf r})\right),\\
&{\bf W}^{ext}_{LS} = b_{4}\nabla \rho ({\bf r}) + b_{4}^{'} \nabla \rho_{q}({\bf r}).
\end{alignat}
\end{subequations}

For ${\cal E}^{ext}_{\mbox{LS}}$ with $b^{'}_{4}= 0$ one gets essentially the same isospin dependence as predicted by relativistic mean field models of Walecka type (see details in ref. \cite{Bender97})

\begin{equation}
  {\bf W}^{\mbox{rmf}}_{LS}=\frac{\hbar^{2}}{[2m - C_{\mbox{eff}}\rho({\bf r})]^{2}}C_{\mbox{eff}}\nabla(\rho({\bf r}).
\end{equation}

The Skyrme parametrization using this choice is called {\bf SKI3} force. As shown in ref. \cite{Rei95}, this force describes correctly the energy shifts in the lead nucleus. On the other hand, one can also vary  $b^{'}_{4}$ while fitting. Hence one has an additional degree of freedom to describe the isoscalar and the isovector channels in the effective potential, like in the other terms of the Skyrme functional. In the {\bf SKI4} force this isospin dependence in the spin-orbit potential in was introduced, see ref. \cite{Rei95}. The parameters  $b_{4}$ and  $b^{'}_{4}$ have been adjusted to the spin-orbit splitting of $^{16}\mbox{O}$ and the isotope shifts of Pb. The fitting has given $b^{'}_{4}\approx -b_{4}$.\\

In the calculation of spin saturated systems, i.e. {\it even-even}-nuclei, the time odd currents are identically zero. In our approach we use a Density-Dependent-Delta-Interaction DDDI) for the usual pairing potential
\begin{equation}
  \upsilon_{\mbox{pair}}({\bf r}-{\bf r}^{'})=V_{0}\left[1-(\frac{\rho({\bf r})}{\rho_{0}})^{\gamma}\cdot \delta({\bf r}-{\bf r}^{'})\right],
\end{equation}
  where $\gamma=1,0$ and $\rho_{0} = 0.16\mbox{ fm}^{-3}$.\\

  Within  our approach we have found that Skyrme forces like SkP or SLy7 do not predict the neutron dripping at all, i.e. the neutron chemical potential remains below $m_{n}c^{2}$ even at rather high densities. These Skyrme forces use ${\cal E}^{\bf J^{2}}_{LS}$ from (25b) as a functional form of the spin-orbit interaction. The most popular Skyrme parametrizations including SkI3, SkI4, SkM*, SIII, SLy4, SLy6 and SLy230a neglect quadratic terms like ${\bf J}^{2}$ or ${\bf J}_{q}^{2}$ and are consistent with the neutron drip-point.

  \subsection{Implementation of electrons}
  In dense stellar matter a crucial role is played by electrons. The interaction between the nucleus and the electron background is very important and must be taken into account. In our Hartree-Fock code this is done by using the method described in ref. \cite{BMG07}. The effects of inhomogeneity in the electron distribution have been studied in ref. \cite{EBM18}. As follows from this analysis, the approximation of uniform density is good enough for the medium-size nuclei considered in this paper. But they may become significant for very big nuclei predicted in the inner crust, see ref. \cite{EBM18}.\\

  The calculations below are done by using the WS method, i.e. the clusterized system is divided into WS cells, each cell containing a nucleus with charge number $Z$ and $Z$ electrons. Therefore the required electrical neutrality of the cell is automatically fulfilled. Using this in the calculation of the neutron star crust means a segmentation of the system into an ensemble of WS cells each containing a  cluster of nuclear matter surrounded by an uniform background of electrons. In contrast to refs. \cite{WS33,WS34} we assume that the size of the WS cell depends only on the baryon density which is simulated. For a fixed baryon density $n_{B}$ and for each nuclear cluster containing Z protons and N neutrons the radius of the Wigner-Seitz cell $R_{WS}$ is calculated from the charge neutrality condition
\begin{equation}
\frac{4\pi}{3}R_{ws}^{3}n_{e}=Z,
\end{equation} 
where $Z$ is the nuclear charge and $n_{e}=\frac{k_{F}^{3}}{3\pi^{2}}$ is the constant background electron density. This gives
\begin{equation}
k_{F}=\big(\,\frac{9\pi Z}{4}\big)\,^{\frac{1}{3}}\cdot\frac{1}{R_{WS}}\approx 1.91\cdot \frac{Z^{1/3}}{R_{WS}},
\end{equation}
where $Z$ is the proton number in the nucleus. The value of $k_{F}$ calculated for every size of the WS cell at fixed baryon density $n_{B}$ determines the Coulomb interaction between the nucleus and the electrons and the energy of the relativistic electron gas.  The radius of the WS cell depends on the mass number of the nucleus $A$ and the fixed baryon density $n_{B}$, 
\begin{equation}
  R_{WS}=\big(\frac{3\cdot A}{4\pi \cdot n_{B}}\big)^{\frac{1}{3}}=(\frac{3}{4\pi})^{\frac{1}{3}}\cdot n^{-\frac{1}{3}}_{B}\cdot A^{\frac{1}{3}}.
\end{equation}  

Inserting this in equation (33) gives a simple formula for the electron Fermi momentum,
\begin{equation}
k_{F}= (3\pi^{2})^{\frac{1}{3}}\cdot n_{B}^{\frac{1}{3}}\cdot \left(\frac{Z}{A}\right)^{\frac{1}{3}}.  
\end{equation}
Instead of baryon density $n_{B}$, below we often use the nucleon mass density $\rho_{B}=m_{N}\times n_{B}$,   
where $m_{N} = \frac{1}{2}(m_{n}+m_{p})=936 \mbox{ MeV}/c^{2}$.\\

The interaction between the electrons and the nucleus in the WS cell is calculated in a self-consistent way, namely, by solving Poisson equation for the electrostatic potential created by both, protons and electrons
\begin{equation}
\bigtriangleup \phi = -e \cdot n_{ch}\equiv -e\cdot(n_{p}-n_{e}).  
\end{equation}
To get a smooth charge distribution the electron density is parametrized with a smooth step-like function $n_{e}(r) = \frac{n_{e}^{0}}{1+exp(\frac{(r-R_{ws})}{a}}$, where $n_{e}^{0}$ is the constant background electron density which ensures charge neutrality in the cell. The diffuseness parameter $a$ is taken to be 0.45 fm, as in ref. \cite{BMG07}. 
The electron density is computed for each individual cell size depending on $n_{B}$ and $(A,Z)$. Finally, after the iterative procedure the electrostatic energy $\frac{1}{2}\int (\nabla \phi)^{2} d^{3}r$ is added to the total energy of the cell.\\ 
The calculational procedure is organized as follows:\\
First, we choose the Skyrme force and start the calculation with a fixed baryon density {\it dens} in units of $[\frac{\mbox{g}}{\mbox{cm}^{3}}]$ . The loop over proton number runs from $Z=2$ up to $Z=130$ and the loop over neutrons runs from $N=Z-2$ to $N=15\cdot Z$, where we take only even Z and N. Then, the Wigner-Seitz cell parameters for every pair (Z, N) are calculated. These are the radius of the cell $R_{WS}$, the Fermi momentum $k_{F}$, the electron chemical potential $\mu_{e}$ and kinetic energy of the ultra-relativistic electron gas, which is
\begin{equation}
E^{kin}_{e}=\frac{3}{4}\cdot k_{F}\cdot Z.    
\end{equation}
Finally, the input files containing $N$, $Z$, $k_{F}$ and $R_{WS}$ are created and the Hartree-Fock program is executed with these input files. The numerical criterion of convergence of iterations to the optimum within the Hartree-Fock approach is chosen to be $10^{-5}$.\\
For Z fixed, the nucleus (N,Z) with the minimal total energy per baryon $E^{min}_{tot}/A$ is selected. An example of such calculations is presented in Fig. 1 for the SkM* force. It shows $E_{tot}/A$ for a set of nuclei (N, Z) with neutron numbers which minimize the total energy calculated at relatively low densities $(1.0 - 9.0)\cdot 10^{6} \frac{\mbox{g}}{\mbox{cm}^{3}}$. The nuclei with optimal proton number Z for several baryon densities lie on the black line. These are ground states predicted at these baryon densities.

\section{Description of the numerical code used for nuclear-structure calculations}

We have performed HFB calculations of nuclear ground states with different Skyrme forces in a broad range of baryon densities. We use the BCS model with delta-like pairing interaction for protons and neutrons. For the pairing strength of neutrons we choose -275,8 MeV and for the pairing strength of protons -291,7 MeV. The number of maximal iterations within the Hartree-Fock procedure is chosen to be 1000 and the relative shift of energy level  $dE$ at the end of the iterations has to be less than $10^{-5}$, as mentioned before. If $\mbox{dE} < 10^{-5}$ is reached, the calculation is considered to be converged. The HFB calculations are done on a grid with the maximal radius equal to the radius of the Wigner-Seitz cell $R_{WS}$. The spacing of the grid points is chosen to be 0,1 fm.\\
As a first step the grid functions defining the size and the fineness of the grid and the corresponding trial wave functions on the grid are constructed. Grid size and grid spacing are taken from the input data. Then the choice of the representation of the baryons and their interaction are done. The nuclear potentials are constructed by using the Skyrme parameters stored in a special file containing all parameter sets. The Skyrme forces used in the code are characterized by the following parameters:
\begin{equation}
\left\{t_{0}, t_{1}, t_{2}, t_{3}, x_{0}, x_{1}, x_{2}, x_{3}, b_{4}, b_{4}', \alpha, \frac{\hbar^{ 2}}{2m}, \mbox{so\_Curr, coul\_Ex, cm\_Corr, dens\_Dep} \right\}.
\end{equation}
Using the baryon data from the input file the wave functions are calculated again with the help of these constructed potentials. With the new wave functions the potentials are calculated self-consistently again.\\
At the end of every iteration loop total energy of the nucleus is calculated as mentioned before. The total kinetic energy is calculated by summing up the kinetic energy and adding the pairing energy for each nucleon represented in a many-body wave function. Then the contributions of the Skyrme mean fields are added. The single-particle wave-functions are weighted with their occupation probabilities and degeneracy factors. The center of mass correction is done at the very last iteration step and normalized with $1/A$.\\
The Hartree-Fock equations during are solved iteratively using the damped gradient step method, see details in refs. \cite{Bender97,LMK91}. The true wave function is found from the iterative equation
\begin{equation}
|\phi_{\alpha}^{n+1}) = {\cal O}\left\{|\phi^{(n)}_{\alpha})- \hat{D}\left[\hat{h}- (\phi_{\alpha}|\hat{h}|\phi_{\alpha})\right]|\phi_{\alpha}^{(n)}) \right\},
\end{equation}
where $\hat{h}$ is the Hamiltonian and only the projections on the diagonal elements using $\Pi^{diag}_ {\alpha}= 1 -|\phi_{\alpha}\rangle\langle \phi_{\alpha}|$ are considered.

The damping operator $\hat{D}$ is defined as
\begin{equation}
  \hat{D}= x_{0}\frac{1}{1 + \frac{\hat{t}}{E_{0}}}
\end{equation}

with the step parameter $x_{0}$, the kinetic energy $\hat{t}$ and the constant energy scale $E_{0}$  which controls the damping. When the iterative process does not conserve the orthonormality of the states , a new set of wave functions has to be generated. The symbol ${\cal O} \{ ... \}$ stands for this orthonormalization procedure after the gradient step.\\
The HFB code was initially  developed in {\it Fortran77} and was ported later to {\it C++} by Bender and Rutz \cite{Bender97,Rutz99}. Then the code was modified as described in ref. \cite{BMG07}, where the uniform electron background was implemented. The code was further developed in ref. \cite{EBM18}, where the neutron gas build by the dripping neutrons was simulated with a special choice of boundary conditions.\\

\section{Predicted Drip point densities and drip point elements}

The drip points are defined as the configurations with a global  minimum of the total energy $E_{tot}/A$ under the condition $\mu_{n}=m_{n}\cdot c^{2}$. It turned out that the results depend essentially on the Skyrme force used in the calculation. For example, using the SKM* parametrization we get the drip point at a baryon density of $4.6\cdot 10^{11}\frac{\mbox{g}}{\mbox{cm}^{3}}$ with $Ti^{82}$ as the drip nucleus. Figure 2 shows the total energy per particle $E_{tot}/A$ as a function of proton number calculated for this density. We have also explored the chemical composition in a region slightly above the neutron drip region. Here {\it free} neutrons appear in the cell. Figure 3 shows the evolution of the nuclear composition with increasing baryon density. We observe that the degree of neutronization increases, i.e. $x=Z/A$ decreases, with increasing baryon density. As mentioned before, we have to exclude some Skyrme forces, like SLy7 or SKP, which do not provide the dripping of neutrons at all. The total neutron number in the cell is determined from the minimization of the total energy $E_{tot}/A$. By performing these calculations one obtains the chemical composition of the neutron drip region as a function of baryon density. Due to the appearance of the neutron gas after the neutron drip in the WS cell one can expect a change of the slope in the EoS of cold nuclear matter, because the lighter neutrons generate an additional pressure (for details see ref. \cite{DouHae01}). \\ 

In the neutron-drip region the strong forces are attractive until the density $1,2\times 10^{14}\frac{\mbox{g}}{\mbox{cm}^{3}}$ is reached,  where the transition to the outer core is expected. Hence the thickness of the inner crust depends strongly on the neutron-drip density. Below we demonstrate precisely how drip-point elements and drip-point densities vary for different Skyrme forces. Comparing the drip-element with the pre-drip element we observe that in every calculation the elements remain the same but the number of dripped neutrons varies from 2 to 12 depending on the Skyrme force used.\\ 
Finally we present the results of our extended calculations carried out for a large variety of Skyrme forces used in the literature. The most popular 7 parametrizations are discussed in some details. Predictions for $\delta$, $x=\frac{Z}{A}$ and $\rho_{ND}$ of other more than 230 parametrizations are presented in the appendix. The Fermi momentum of the electron gas $k_{F}$ at the drip point where the slope of the EoS changes can be calculated using equation (35). With this method only $x=Z/A$ can be obtained, the special drip element can not be determined. All the values for $x$ with parameters taken from \cite{DuSto12} are in the same order: $x \sim 0.3$.\\
\\
In this section we present our calculations for the 7 most popular Skyrme forces: SkM*, SkI3, SkI4,SIII, SLy4, SLy6 and the newer SLy230a.\\

  \subsection{Skyrme force $\mbox{SKM}^{*}$}
  The previously used Skyrme force SkM has been extensively studied for both spherical and deformed nuclei through Hartree-Fock plus BCS calculations \cite{SkM}. Ground-state radii and multipole moments are found in excellent agreement with experimental data. But nevertheless binding energies were systematically too high and fission barriers were significantly too low. The modified Skyrme force $\mbox{SkM}^{*}$ was the first Skyrme force with reasonable incompressibility as well as better fission properties \cite{SkM_star}. Neutron and proton single-particle energies in the $_{40}^{90}Zr$, $_{82}^{208}Pb$ and $_{94}^{240}Pu$ were computed and compared with experimental energies taken from the compilation of ref. \cite{WaBo77}. With SkM* it was possible to calculate the fission barrier of $_{94}^{240}Pu$ rather accurately. This Skyrme force uses ${\cal E}^{\mbox std}_{LS}$ for the spin-orbit interaction. Calculations with SKM* yield $_{22}^{80}Ti$ as the pre-drip element at a density of $4,5 \cdot 10^{11} \frac{\mbox{g}}{\mbox{cm}^{3}}$. The neutron drip occurs at a density of $4,6 \cdot 10^{11} \frac{\mbox{g}}{\mbox{cm}^{3}}$ starting with the Titanium isotope $_{22}^{84}Ti$, which corresponds to $x=0.268$. Hence the neutron gas in the cell consists of 2 neutrons in a spherical WS-cell with radius of 39,96 fm. For SKM* there exists no magic neutron number of the pre-drip element. But this Skyrme parametrization favors the N=50 shell in the transition region from outer to inner crust. The drip density calculated with SKM* is identical to the commonly accepted value  $4,6 \cdot 10^{11} \frac{\mbox{g}}{\mbox{cm}^{3}}$ obtained in \cite{BBP71}. The evolution of the ground state in the drip region is illustrated in Fig. 3. One can see that during the neutron drip the element remains the same, but with increasing density the neutronization increases, i.e. $x=Z/A$ decreases. With the estimate based on the symmetry energy described before the drip occurs at a density of $4,274\cdot 10^{11} \frac{\mbox{g}}{\mbox{cm}^{3}}$ with $x=0.383$ (see chapter II).   
 \\ \\
{\begin{tabular}{|c|c|c|c|c|c|c|c|}\hline
Skyrme force &$\rho_{pre-drip}$ $[\frac{\mbox{g}}{\mbox{cm}^{3}}]$ & pre-drip nuclei & drip element&$\rho_{drip}$ $[\frac{\mbox{g}}{\mbox{cm}^{3}}]$&$\frac{E_{tot}}{A}$[MeV]& $x=\frac{Z}{A}$ &$\mu_{n}$[MeV]\\ \hline\hline
$\mbox{SkM}^{*}$&$4,5 \cdot 10^{11}$&$_{22}^{80}Ti$ & $_{22}^{82}Ti$&$4,6\cdot 10^{11}$&-1,3540&0.268 &$2,01 \cdot 10^{-1}$\\ \hline
  \end{tabular}
  \\ \\
  
\subsection{Skyrme forces SkI3 and SkI4}

The SkIx Skyrme forces are based on calculations performed by P.-G. Reinhard and H. Flocard in 1995 \cite{Rei95}. They used least-square fit of nuclear ground-state properties \cite{Fri86} and took experimental data of exotic nuclei into account. The Skyrme force SkI1 has the standard spin-orbit coupling ${\cal E}^{\mbox std}_{LS}$ and $b_{4}^{'} = b_{4}$. This Skyrme force does not provide the dripping of neutrons and therefore is not considered here. The SkI3 force has a generalized spin-orbit coupling which is the non-relativistic limit of a relativistic mean-field model (${\cal E}^{\mbox ext}_{LS}$ and $b^{ '}_{4}=0$). The Skyrme force SkI4 similarly to SkI3 uses a generalized spin-orbit coupling ${\cal E}^{\mbox ext}_{LS}$. The pre-drip element calculated with SkI3 is $_{34}^{116}Se$ with a {\bf magic N=82 shell} at a density of $3,5 \cdot 10^{11} \frac{\mbox{g}}{\mbox{cm}^{3}}$. The drip element is the Selenium isotope $_{34}^{118}Se$ at a density of $3,6\cdot 10^{11} \frac{\mbox{g}}{\mbox{cm}^{3}}$ with $x=0.288$. The neutron gas at the drip point consists of 2 neutrons in a spherical cell with radius 48.96 fm. Using tihe Skyrme force SkI4 one gets as pre-drip element $_{28}^{98}Ni$ with the magic proton shell $P=28$ at a slightly higher density of $3.7 \cdot 10^{11} \frac{\mbox{g}}{\mbox{cm}^{3}}$. The drip element is the Nickel isotope $_{28}^{106}Ni$ with $x=0.264$ at a density of $3,8\cdot 10^{11} \frac{\mbox{g}}{\mbox{cm}^{3}}$. The neutron gas at the drip consists of 8 neutrons in a spherical cell with radius 46,39 fm. In contrast the the SkM* force the neutron drip densities of SkI3 ans SkI4 are both lower than predicted in \cite{BPS71}. With the estimates based on the symmetry energy the drip occurs at a density of $4,46\cdot 10^{11} \frac{\mbox{g}}{\mbox{cm}^{3}}$ with $x=0.396$ using SKI3 and  $4,42\cdot 10^{11} \frac{\mbox{g}}{\mbox{cm}^{3}}$ with $x=0.379$ using SKI4.\\
\\ \\  
\begin{tabular}{|c|c|c|c|c|c|c|c|}\hline
Skyrme force &$\rho_{pre-drip}$ $[\frac{\mbox{g}}{\mbox{cm}^{3}}]$ & pre-drip nuclei & drip element&$\rho_{drip}$ $[\frac{\mbox{g}}{\mbox{cm}^{3}}]$&$\frac{E_{tot}}{A}$[MeV] & $x=\frac{Z}{A}$ &$\mu_{n}$[MeV]\\ \hline\hline
SkI3&$3,5 \cdot 10^{11}$&$_{34}^{116}Se$ & $_{34}^{118}Se$&$3,6\cdot 10^{11}$&-1,6701&0.288 &$2,73 \cdot 10^{-1}$\\ \hline
SkI4&$3,7 \cdot 10^{11}$&$_{28}^{98}Ni$&$_{28}^{106}Ni$&$3,8 \cdot 10^{11}$&-1,538&0.264&$1,08\cdot 10^{-1}$\\ \hline
\end{tabular}
\\ \\
\subsection{Skyrme force SIII}

The Skyrme force SIII was proposed by Beiner et al. in 1975 \cite{SIII}. It is one of the oldest Skyrme forces still used today, which uses ${\cal E}^{\mbox std}_{LS}$ for the spin-orbit coupling. The authors have performed a detailed study of the influence of the force parameters on the binding energies, charge densities, radii and single-particle energies. They explored also magic nuclei. For the SIII force the pre-drip element is $_{36}^{118}Kr$ at  $\rho=3,6\cdot 10^{11} \frac{\mbox{g}}{\mbox{cm}^{3}}$ with {\bf magic N=82 shell}. The neutron drip starts at a density of $\rho=3,7\cdot 10^{11} \frac{\mbox{g}}{\mbox{cm}^{3}}$ with the Krypton isotope $_{36}^{130}Kr$ and $x=0.277$, which is even lower than the neutron drip density predicted in \cite{BBP71}. Here the relatively large number of 12 neutrons form the neutron gas in a spherical cell with radius of 50,10 fm. With the estimates based on the symmetry energy the drip occurs at a density of $4,34\cdot 10^{11} \frac{\mbox{g}}{\mbox{cm}^{3}}$ with $x=0.375$.\\
\\
{\begin{tabular}{|c|c|c|c|c|c|c|c|}\hline
Skyrme force &$\rho_{pre-drip}$ $[\frac{\mbox{g}}{\mbox{cm}^{3}}]$ & pre-drip nuclei & drip element&$\rho_{drip}$ $[\frac{\mbox{g}}{\mbox{cm}^{3}}]$&$\frac{E_{tot}}{A}$[MeV] & $x=\frac{Z}{A}$ &$\mu_{n}$[MeV]\\ \hline\hline
SIII&$3,6 \cdot 10^{11}$&$_{36}^{118}Kr$ & $_{36}^{130}Kr$&$3,7\cdot 10^{11}$&-1,5468&0.277 &$2,48 \cdot 10^{-1}$\\ \hline
\end{tabular}
\\ \\
 \subsection{Skyrme forces SLy4 and SLy6}

 The construction of the SLyx Skyrme forces was motivated by the most accurate description of neutron-rich nuclei. The SLy6 force does not use the ${\bf J}^{2}$-term. The correction due to the center of mass motion is introduced in a self-consistent way using the full microscopic ansatz (details see in ref. \cite{SLy}). The neutron drip starts at a density of $\rho=3,6\cdot 10^{11} \frac{\mbox{g}}{\mbox{cm}^{3}}$ with nucleus  $_{36}^{124}Kr$ and $x=0.290$. The pre-drip isotope $_{36}^{118}Kr$ with a {\bf magic N=82 shell} appears at a density of $\rho=3,5 \cdot 10^{11} \frac{\mbox{g}}{\mbox{cm}^{3}}$. Hence the neutron gas at the drip consists of 6 unbound neutrons in the spherical cell with a radius of $49,77$ fm.\\
The Skyrme force SLy4 uses a simple version for the center of mass correction where only diagonal elements of
\begin{equation}
\langle \hat{{\bf P}}^{2}_{cm}\rangle=\sum_{k,k' \geq 0}\sum_{m,m'\geq 0}{\bf p}_{k,k'}{\bf p}_{m,m'}\langle \mbox{BCS}|\hat{a}^{\dagger}_{k}\hat{a}_{k'}\hat{a}^{\dagger}_{m}\hat{a}_{m'}|\mbox{BCS}\rangle   
\end{equation}   
are taken into account, for details see ref. \cite{Bender97}. The pre-drip element is $_{38}^{120}Sr$ with a {\bf magic N=82 shell}. The neutron gas at the drip point consists of 4 neutrons in a spherical shell of radius $50,73$ fm. The neutron drip starts at a density of $\rho=3,4\cdot 10^{11} \frac{\mbox{g}}{\mbox{cm}^{3}}$ with drip element $_{38}^{124}Sr$ and $x=0.306$.\\
The parametrizations SLy4 and Sly6 are very similar. Most predictions such as compressibility $K_{\infty}$, sum-rule enhancement factor $\kappa$  or the asymmetry coefficient $a_{sym}$ are identical. There are only some differences in (a) the energy per nucleon (15,97 MeV for SLy4 and 15,90 MeV for SLy6), (b) the nuclear density $n_{0}$ (0.160 $\mbox{ fm}^{-3}$ for SLy4 and 0,159 $\mbox{ fm}^{-3}$ for SLy6), (c) the effective mass $m^{*}/m$ (0,695 for SLy4 and 0,690 for SLy6) and (d) the use of the spin-orbit coupling: Sly4 uses ${\cal E}^{\mbox{ext}}_{LS}$, whereas Sly6 uses ${\cal E}^{\mbox{std}}_{LS}$. We found that SLy6 yields a slightly higher neutron drip density with $_{36}^{124}Kr$ as drip-element. But actually the two drip elements of SLy4 and SLy6 are isobars. One more force of this family, SLy7, is using ${\cal E}^{({\bf J}^{2})}_{LS}$ for the spin-orbit interaction which we do not include in our analysis. With the estimates based on the symmetry energy the drip occurs at a density of $4,45\cdot 10^{11} \frac{\mbox{g}}{\mbox{cm}^{3}}$ with $x=0.388$ using SLy4 and  $4,40\cdot 10^{11} \frac{\mbox{g}}{\mbox{cm}^{3}}$ with the same value $x=0.388$ using SLy6.\\
\\  
{\begin{tabular}{|c|c|c|c|c|c|c|c|}\hline
    Skyrme force &$\rho_{pre-drip}$ $[\frac{\mbox{g}}{\mbox{cm}^{3}}]$ & pre-drip nuclei & drip element&$\rho_{drip}$ $[\frac{\mbox{g}}{\mbox{cm}^{3}}]$&$\frac{E_{tot}}{A}$[MeV] & $x=\frac{Z}{A}$ &$\mu_{n}$[MeV]\\ \hline\hline
SLy4&$3,3 \cdot 10^{11}$&$_{38}^{120}Sr$ & $_{38}^{124}Sr$&$3,4\cdot 10^{11}$&-1,6490&0.306&$2,27 \cdot 10^{-1}$\\ \hline
SLy6&$3,5 \cdot 10^{11}$&$_{36}^{118}Kr$ & $_{36}^{124}Kr$&$3,6\cdot 10^{11}$&-1,521258&0.29 &$3,87 \cdot 10^{-1}$\\ \hline
  \end{tabular}
  \\  \\

 \subsection{Skyrme force SLy230a}

The force SLy230a \cite{SLy} is using the following parameters for symmetric nuclear matter:\\


\begin{tabular}{|c|c|c|c|c|} \hline
E/A [Mev]&$\rho_{0} \frac{1}{\mbox{\mbox{fm}}^{3}}$&K[MeV]&$E_{sym}$&$m^{*}/m$\\\hline \hline
-15.988&0.160&229.87&31.97&0.697\\ \hline
\end{tabular}
  \\ \\
  This parametrization is quite common and matches very well to the binding energies and the charge radii of doubly-magic nuclei. Pre-drip element at a density of $\rho=2,8 \cdot 10^{11} \frac{\mbox{g}}{\mbox{cm}^{3}}$ is $_{42}^{124}Mo$ with a {\bf magic N=82 shell}. In ref. \cite{NegVauth73} the authors found Molybdenum isotope to be the last pre-drip nucleus. Actually they found the sequence of pre-drip nuclei to be  $ _{42}^{124}\mbox{Mo}$ - $_{40}^{122}\mbox{Zr}$ - $_{38}^{120}\mbox{Sr}$ - $_{36}^{118}\mbox{Kr}$. The neutron gas at the drip point with drip element $_{42}^{126}Mo$ and $x=0.333$ is very diluted with 2 neutrons in a spherical cell of radius 53,784 fm.  With the estimates based on the symmetry energy the drip occurs at a density of $4,46\cdot 10^{11} \frac{\mbox{g}}{\mbox{cm}^{3}}$ with $x=0.388$ as SLy4 or SLy6.\\
\\\\
\\
{\begin{tabular}{|c|c|c|c|c|c|c|c|}\hline
Skyrme force &$\rho_{pre-drip}$ $[\frac{\mbox{\mbox{g}}}{\mbox{cm}^{3}}]$ & pre-drip nuclei & drip element&$\rho_{drip}$ $[\frac{\mbox{g}}{\mbox{cm}^{3}}]$&$\frac{E_{tot}}{A}$[MeV] & $x=\frac{Z}{A}$ &$\mu_{n}$ [MeV]\\ \hline\hline
SLy230a&$2,8 \cdot 10^{11}$&$_{42}^{124}Mo$ & $_{42}^{126}Mo$&$2,9\cdot 10^{11}$&-1,6788&0.333 &$2,044 \cdot 10^{-1}$\\ \hline
\end{tabular}
\\

  The most important characteristics of the drip-point for the 7 selected Skyrme parametrizations are summarized in the table below. One can see that the number of dripped neutrons in the cell varies from 2 to 12 and the neutron-gas has densities around $n_{n}\sim 10^{-5} \mbox{fm}^{-3}$, but the chemical potentials of the dripped neutrons are almost of the same value: $\mu_{n} \sim (0,2-0,3)$ MeV. Although the proton numbers of the drip-point elements vary from 22 to 42 the chemical potential of the electrons $\mu_{e}$ has also nearly the same value for these Skyrme forces, around {\bf 25 MeV}.\\
Therefore the lower drip density is the most important point of our investigation. Obviously, for a lower drip point density the outer crust would be thinner. This may significantly change the observable signatures of neutron stars as pointed out e.g. in \cite{Glend07}.  \\
\\
{\bf Drip-point characteristics for 7 most popular Skyrme forces}\\
\\
 {\begin{tabular}{|c|c|c|c|c|c|c|c|c|}\hline
     Skyrme&drip element&$n_{ND}$ $[\frac{\mbox{g}}{\mbox{cm}^{3}}]$ &n-gas&$\langle r\rangle_{\mbox{bar}}$[fm]&$ \langle r\rangle_{WS}$[fm]&$n_{n}$[$\mbox{fm}^{-3}$] &$\mu_{n}-m_{n}c^{2}$[MeV]&$\mu_{e}$[MeV]\\ \hline\hline

     SkI3&$_{34}^{118}Se$ & $3,6\cdot 10^{11}$ & 2 & 4,84 & 48,96 &$5,56\cdot 10^{-6}$& $2,73\cdot 10^{-1}$&25,07\\ \hline

     SkI4&$_{28}^{106}Ni$ &$3,8 \cdot 10^{11}$ & 8  & 4,66& 46,39 & $2,63\cdot10^{-5}$& $1,08 \cdot 10^{-1}$&24,78\\ \hline

     $\mbox{SkM}^{*}$&$_{22}^{82}Ti$& $4,6 \cdot 10^{11}$&2&4,45&39,96&$1,07\cdot 10^{-5}$ &$2,01 \cdot 10^{-1}$&25,55\\ \hline
     
        SIII        &$_{36}^{130}Kr$&$3,7\cdot 10^{11}$&12&4,745&50,10&$3,07\cdot10^{-5}$  &$2,48 \cdot 10^{-1}$&24,95\\ \hline

      SLy4&          $_{38}^{124}Sr$&$3,4\cdot 10^{11}$&4&4,76&50,73&$9,83\cdot 10^{-6}$&$2,27\cdot 10^{-1}$&25,09\\ \hline

      SLy6&$_{36}^{124}Kr$ & $3,6\cdot 10^{11}$&6&4,77&49,77 &$1,57\cdot  10^{-5}$&$3,87 \cdot 10^{-1}$&25,12\\ \hline

      SLy230a& $_{42}^{126}Mo$&$2,9\cdot 10^{11}$&2&4,79&53,78&$3,82\cdot10^{-6}$ &$2,04 \cdot 10^{-1}$&24,47\\ \hline
   \end{tabular}
\\
\section{Discussion and conclusions}

The presented calculations demonstrate that almost all  obtained drip densities are lower than the commonly accepted value of Baym, Bethe and Pethick, ref. \cite{BBP71}. Only the $\mbox{SKM}^{*}$-force matches quite well this prediction. But on the other hand, the Wigner-Seitz cell is mainly filled by the relativistic electron gas, only about $\frac{1}{1000}$ of the volume of the WS-cell is occupied by the nucleus. Therefore, practically the whole pressure comes from the electrons. In our calculations the interaction of the nucleus with the surrounding electron gas, the so-called lattice energy, is taken into account in a new self-consistent way. At every baryon density we have calculated every single nucleus and obtained precise values of the neutron chemical potential and the neutron gas density in the WS cell. These observables are calculated within a fully microscopic approach and directly linked to the parameters of the special Skyrme force. Within our approach we are able to describe exactly the nuclear changes with increasing baryon density.  So, as a main result of our investigations, most of our drip densities are lower than the commonly accepted values predicted between 4 and 5 $\times 10^{11} \frac{\mbox{g}}{\mbox{cm}^{3}}$ \cite{WeberF}. The first calculations of Baym, Pethik and Sutherland  \cite{BPS71}  predicted the neutron drip density $4,3 \cdot10^{11} \frac{\mbox{g}}{\mbox{cm}^{3}}$. They found the electron chemical potential to be $26.2$ MeV, which is only 1 MeV higher than in our calculations, and they determined $_{36}^{118}Kr$ as the drip element. In their calculation {\it Kr} remains the favoured element even somewhat beyond that drip-point. Using a compressible liquid-drop model Baym, Bethe and Pethik \cite{BBP71} confirmed the value $4,3\cdot 10^{11}\frac{\mbox{g}}{\mbox{cm}^{3}}$ as the density where the neutron-drip starts. But even the drip densities based on the estimates using the symmetry energy of ref. \cite{DuSto12} do suggest lower densities of the neutron drip for most forces used (see appendix).\\
However, at such high densities more accurate calculations of electron-nucleus interaction are required. Such calculations have been carried out in ref. \cite{EBM18} where the electron distributions were obtained by solving the Poisson equation for electrostatic potential in combination with nuclear structure calculations using a RMF model.\\
Magic numbers play an important role in nuclear structures and stability. The pre-drip elements, most of them stabilized by magic neutron shells,  are the last stable neutron-rich elements before neutrons drip out. Hence one can expect that magic neutron shells may stabilize the configurations even at higher densities. Negele and Vautherin found $_{36}^{118}Kr$ to be the last pre-drip element, stabilized by the {\bf magic N=82 shell} \cite{NegVauth73}. This magic shell appears also in the Skyrme parametrizations SKI3, SIII, SLy4, SLy6 and SLy230a. On the other hand, the Skyrme force SKM* belongs to the group of parametrizations favouring the {\bf magic N=50 shell} in the drip region. This shell survives until the density $\rho = 7,0\cdot 10^{10} \frac{\mbox{g}}{\mbox{cm}^{3}}$. At higher densities the number of neutrons increases up to 60 for the drip element $_{22}^{82}Ti$ at $\rho=4,6\cdot 10^{11} \frac{\mbox{g}}{\mbox{cm}^{3}}$. According to ref. \cite{Roco-Maza12}, there are mainly two types of parametrizations, favouring either the {\bf N=50} shell or the {\bf N=82} shell in the neutron-drip region. It is also shown in ref. \cite{Roco-Maza12}, that the nuclear symmetry energy has a great influence on the chemical composition of the outer crust and therefore on the magic neutron shells in the drip-region. In Fig. 2 one can observe that there is a local minimum at proton number $Z=50$. But this minimum gets weaker with increasing density and is shifted to a proton number smaller than 50. Hence in our calculations the magic P=50 shell is {\bf quenched} at the edge of the outer crust. \\
Looking at the Skyrme forces used in our analysis, one can see that the values of $E/A$ of the drip elements vary between $-1.6$ Mev and $-1.5$ MeV, i.e. are rather close to each other. Only for SkM* the $E/A$ value is slightly higher ($-1.35$ MeV). This can be explained by the smaller neutron number of the magic shell appearing in the drip region. Accordingly,  the drip element $_{22} ^{82}Ti $ obtained with SkM* has the smallest proton number ($Z=22$) of all Skyrme forces within our investigation. But on the other hand, the atomic numbers of the drip elements obtained for other forces vary from $28$ (Ni) for SkI4 up to $42$ (Mo) for SLy230a. The element $\ce{^{124}_{42}Mo}$ appears also in the calculations of Negele-Vautherin, ref. \cite{NegVauth73}, but as a pre-drip element. Looking at the mass-energy density profiles of neutron stars in ref. \cite{Glend00}, one may conclude that the neutron drip-density is about $\rho \sim 4\cdot 10^{11} \frac{\mbox{g}}{\mbox{cm}^{3}}$. This is a slightly lower than the value reported in ref. \cite{BBP71}. Using our method most drip point densities are lower than 4 $\cdot 10^{11} \frac{\mbox{g}}{\mbox{cm}^{3}}$. Applying the method described in \cite{HaePoYa07} to all Skyrme forces listed in \cite{DuSto12} we have obtained the table presented in Appendix A which includes more than 200 Skyrme forces. As one can see, drip densities vary from $ 2,9 \cdot 10^{11} \frac{\mbox{g}}{\mbox{cm}^{3}}$ to $4,6 \cdot 10^{11}\frac{\mbox{g}}{\mbox{cm}^{3}}$. Obviously, a lower drip density corresponds to a thinner outer crust. These results may be useful for calculating cooling rates and mechanical properties of the outer crust. 
\\ \\
Finally, we conclude with a few remarks:\\
\\
First, one can see that due to $\rho_{ND}$ the modern Skyrme parametrization $\mbox{SKM}^{*}$ ($\rho_{ND} = 4,6 \cdot 10^{11} \frac{\mbox{g}}{\mbox{cm}^{3}}$) is indeed better parametrization than the older SKM force ($\rho_{ND}=4,3 \cdot 10^{11}\frac{\mbox{g}}{\mbox{cm}^{3}}$), supposing that the neutron drip density predicted by \cite{BBP71} is true.\\
\\
Second, the drip densities obtained with SkTK, $\mbox{SIII}^{*}$ and the family ZRXx are too high and\\
third, the drip densities obtained with the family v070 - v110 are too low.\\
\\
Third, the Skyrme forces ZR3a, ZR3b and ZR3c with $S_{0}<0$ are very special and have been excluded from our analysis.\\
\\
We believe that our calculations will be useful for further studies of neutron-star physics, including merger events, cooling rates, crust characteristics, nucleosynthesis etc. Our analysis can be easily extended to other Skyrme forces to describe nuclear systems in dense neutron-rich environments.\\
\\
The authors thank T. Buervenich and D. Blaschke for fruitful discussions, and the FIAS IT group for providing computational resources and continuous support.

\newpage
\appendix
\section{Neutron drip-point characteristics for different Skyrme forces}
{\begin{tabular}{|l|c|c|c|c|c|}\hline
Skyrme force &$E_{0}$ &$ S_{0}$ &$ \delta$ & x &$ \rho_{ND}[\frac{\mbox{g}}{\mbox{cm}^{3}}]$ \\ \hline\hline
BSK1&-15.81&27.81&0.2524&0.3738&$4.31\cdot 10^{11}$ \\ \hline 
BSK2&-15.80&28.00&0.2507&0.3746&$4.30\cdot 10^{11}$ \\ \hline 
BSK2'&-15.79&28.00&0.2506&0.3747&$4.29\cdot 10^{11}$ \\ \hline 
BSK3&-15.81&27.93&0.2514&0.3743&$4.31\cdot 10^{11}$ \\ \hline 
BSK4&-15.77&28.00&0.2503&0.3748&$4.28\cdot 10^{11}$ \\ \hline 
BSK5&-15.80&28.70&0.2452&0.3774&$4.30\cdot 10^{11}$ \\ \hline
BSK6&-15.75&28.00&0.25&0.375&$4.26\cdot 10^{11}$ \\ \hline
BSK7&-15.76&28.00&0.2501&0.3749&$4.27\cdot 10^{11}$ \\ \hline
BSK8&-15.83&28.00&0.2511&0.3744&$4.32\cdot 10^{11}$ \\ \hline
BSK9&-15.92&30.00&0.2372&0.3814&$4.40\cdot 10^{11}$ \\ \hline
BSK10&-15.91&30.00&0.2371&0.3814&$4.39\cdot 10^{11}$ \\ \hline
BSK11&-15.86&30.00&0.2346&0.3818&$4.35\cdot 10^{11}$ \\ \hline
BSK12&-15.86&30.00&0.2364&0.3818&$4.35\cdot 10^{11}$ \\ \hline
BSK13&-15.86&30.00&0.2364&0.3818&$4.35\cdot 10^{11}$ \\ \hline
BSK14&-15.85&30.00&0.2363&0.3818&$4.34\cdot 10^{11}$ \\ \hline
BSK15&-16.04&30.00&0.2388&0.3806&$4.50\cdot 10^{11}$ \\ \hline
BSK16&-16.05&30.00&0.2389&0.3805&$4.50\cdot 10^{11}$ \\ \hline
BSK17&-16.06&30.00&0.2390&0.3805&$4.50\cdot 10^{11}$ \\ \hline
BSK18&-16.06&30.00&0.2390&0.3805&$4.50\cdot 10^{11}$ \\ \hline
BSK19&-16.08&30.00&0.2393&0.3803&$4.53\cdot 10^{11}$ \\ \hline
BSK20&-16.08&30.00&0.2393&0.3803&$4.53\cdot 10^{11}$ \\ \hline
BSK21&-16.05&30.00&0.2389&0.3805&$4.50\cdot 10^{11}$ \\ \hline
E&-16.13&27.66&0.2582&0.3709&$4.57\cdot 10^{11}$ \\ \hline
Es&-16.02&27.44&0.2599&0.37&$4.56\cdot 10^{11}$ \\ \hline
$\mbox{f}_{-}$&-16.02&32.00&0.225&0.38750&$4.49\cdot 10^{11}$ \\ \hline
$\mbox{f}_{+}$&-16.04&32.00&0.2252&0.3874&$4.50\cdot 10^{11}$ \\ \hline
$\mbox{f}_{0}$&-16.03&32.00&0.2251&0.3874&$4.49\cdot 10^{11}$ \\ \hline
FPLyon&-15.92&30.93&0.2307&0.3846&$4.40\cdot 10^{11}$ \\ \hline
Gs&-15.59&31.13&0.2307&0.3846&$4.49\cdot 10^{11}$ \\ \hline
GS1&-16.03&28.86&0.2471&0.3764&$4.49\cdot 10^{11}$ \\ \hline
\end{tabular}
\newpage
{\begin{tabular}{|l|c|c|c|c|c|}\hline
Skyrme force &$E_{0}$ &$ S_{0}$ &$ \delta$ & x &$ \rho_{ND}[\frac{\mbox{g}}{\mbox{cm}^{3}}]$ \\ \hline\hline
GS2&-16.01&25.96&0.2715&0.3642&$4.48\cdot 10^{11}$ \\ \hline
GS3&-16.00&21.49&0.3208&0.3396&$4.49\cdot 10^{11}$ \\ \hline
GS4&-15.96&12.83&0.4980&0.251&$4.84\cdot 10^{11}$ \\ \hline
GS5&-15.91&18.70&0.3604&0.3198&$4.46\cdot 10^{11}$ \\ \hline
GS6&-16.04&14.33&0.4558&0.2721&$4.77\cdot 10^{11}$ \\ \hline
GSkI&-16.02&32.03&0.2248&0.3876&$4.49\cdot 10^{11}$ \\ \hline
GSkII&-16.12&30.49&0.2364&0.3818&$4.57\cdot 10^{11}$ \\ \hline
KDE&-15.99&31.97&0.2248&0.3876&$4.46\cdot 10^{11}$ \\ \hline
KDE0v&-16.10&32.98&0.2199&0.39&$4.56\cdot 10^{11}$ \\ \hline
KDE0v1&-16.23&34.58&0.2122&0.3939&$4.67\cdot 10^{11}$ \\ \hline
LNS&-15.32&33.43&0.2076&0.3962&$3.93\cdot 10^{11}$ \\ \hline
MSk1&-15.83&30.00&0.2360&0.382&$4.33\cdot 10^{11}$ \\ \hline
MSk2&-15.83&30.00&0.2360&0.382&$4.33\cdot 10^{11}$ \\ \hline
MSk3&-15.79&28.00&0.2506&0.3747&$4.29\cdot 10^{11}$ \\ \hline
MSk4&-15.79&28.00&0.2506&0.3747&$4.29\cdot 10^{11}$ \\ \hline
MSk5&-15.79&28.00&0.2506&0.3747&$4.29\cdot 10^{11}$ \\ \hline
$\mbox{MSk5}^{*}$&-15.78&28.00&0.2504&0.3748&$4.28\cdot 10^{11}$ \\ \hline
MSk6&-15.79&28.00&0.2506&0.3747&$4.29\cdot 10^{11}$ \\ \hline
MSk7&-15.80&27.95&0.2511&0.3744&$4.30\cdot 10^{11}$ \\ \hline
MSk8&-15.80&27.93&0.2513&0.3743&$4.30\cdot 10^{11}$ \\ \hline
MSk9&-15.80&28.00&0.002507&0.3746&$4.30\cdot 10^{11}$ \\ \hline
MSkA&-15.99&30.35&0.2356&0.3822&$4.45\cdot 10^{11}$ \\ \hline
MSL0&-16.00&30.00&0.002383&0.03808&$4.47\cdot 10^{11}$ \\ \hline
NRAPR&-15.85&32.78&0.2180&0.391&$4.35\cdot 10^{11}$ \\ \hline
PRC45&-15.82&51.01&0.1446&0.4277&$4.37\cdot 10^{11}$ \\ \hline
RATP&-16.05&29.26&0.2444&0.03778&$4.51\cdot 10^{11}$ \\ \hline
Rs&-15.59&30.82&0.2271&0.3864&$4.13\cdot 10^{11}$ \\ \hline
Sefm068&-15.92&88.57&0.0862&0.4569&$4.0\cdot 10^{11}$ \\ \hline
Sefm074&-15.81&33.40&0.2138&0.3931&$4.31\cdot 10^{11}$ \\ \hline
Sefm081&-15.69&30.76&0.2288&0.3856&$4.21\cdot 10^{11}$ \\ \hline
\end{tabular}
\newpage
\begin{tabular}{|l|c|c|c|c|c|}\hline
Skyrme force &$E_{0}$ &$ S_{0}$ &$ \delta$ & x &$ \rho_{ND}[\frac{\mbox{g}}{\mbox{cm}^{3}}]$ \\ \hline\hline  
Sefm09&-15.55&27.78&0.2489&0.3755&$4.10\cdot 10^{11}$ \\ \hline
Sefm1&-15.40&24.81&0.2731&0.3634&$3.99\cdot 10^{11}$ \\ \hline
SGI&-15.89&28.33&0.002493&0.3753&$4.37\cdot 10^{11}$ \\ \hline
SGII&-15.60&26.83&0.2575&0.37125&$4.14\cdot 10^{11}$ \\ \hline
SGOI&-16.63&45.20&0.1696&0.4152&$5.05\cdot 10^{11}$ \\ \hline
SGOII&-16.70&93.98&0.0852&0.4574&$5.23\cdot 10^{11}$ \\ \hline
SI&-15.99&29.24&0.2437&0.3781&$4.46\cdot 10^{11}$ \\ \hline
SII&-15.99&34.16&0.2116&0.3942&$4.46\cdot 10^{11}$ \\ \hline
SIII&-15.85&28.16&0.2501&0.3749&$4.34\cdot 10^{11}$ \\ \hline
$\mbox{SIII}^{*}$&-16.07&31.97&0.2583&0.3708&${\bf 7.07\cdot 10^{11}}$ \\ \hline
SIV&-15.96&31.22&0.2293&0.3853&$4.43\cdot 10^{11}$ \\ \hline
Sk1'&-15.99&29.35&0.2429&0.3785&$4.46\cdot 10^{11}$ \\ \hline  
SK255&-16.33&37.40&0.1986&0.4007&$4.76\cdot 10^{11}$ \\ \hline
SK272&-16.28&37.40&0.1980&0.401&$4.72\cdot 10^{11}$ \\ \hline
SkA&-15.99&32.91&0.19&0.3905&$4.30\cdot 10^{11}$ \\ \hline
SkA25s20&-16.07&33.78&0.2148&0.3926&$4.53\cdot 10^{11}$ \\ \hline
SkA35s15&-16.01&30.56&0.446&0.3827&$4.77\cdot 10^{11}$ \\ \hline
SkA35s20&-16.08&33.57&0.2164&0.3918&$4.56\cdot 10^{11}$ \\ \hline
SkA35s25&-16.14&36.98&0.1985&0.4007&$4.6\cdot 10^{11}$ \\ \hline
SkA45s20&-16.08&33.39&0.2172&0.3914&$4.54\cdot 10^{11}$ \\ \hline
SkB&-15.99&23.88&0.2921&0.3539&$4.47\cdot 10^{11}$ \\ \hline
SkI1&-15.95&37.53&0.1937&0.4031&$4.44\cdot 10^{11}$ \\ \hline
SkI2&-15.78&33.37&0.2136&0.3922&$4.37\cdot 10^{11}$ \\ \hline
SkI3&-15.98&34.83&0.2078&0.3961&$4.46\cdot 10^{11}$ \\ \hline
SkI4&-15.95&29.50&0.2412&0.3794&$4.42\cdot 10^{11}$ \\ \hline
SkI5&-15.85&36.64&0.1969&0.4015&$4.36\cdot 10^{11}$ \\ \hline
SkI6&-15.89&29.90&0.2375&0.3812&$4.38\cdot 10^{11}$ \\ \hline
SkM&-15.77&30.75&0.2230&0.385&${\bf 3.90\cdot 10^{11}}$ \\ \hline
$\mbox{SkM}^{*}$&-15.77&30.03&0.2349&0.38255&$4.274\cdot 10^{11}$ \\ \hline
SkM1&-15.77&25.17&0.2753&0.3623&$4.28\cdot 10^{11}$ \\ \hline
SkMP&-15.56&29.89&0.2331&0.3834&$4.11\cdot 10^{11}$ \\ \hline
\end{tabular}
\newpage
{\begin{tabular}{|l|c|c|c|c|c|}\hline
Skyrme force &$E_{0}$ &$ S_{0}$ &$ \delta$ & x &$\rho_{ND}[\frac{\mbox{g}}{\mbox{cm}^{3}}]$ \\ \hline\hline     
SkO&-15.84&31.97&0.2289&0.3855&$4.73\cdot 10^{11}$ \\ \hline
SkO'&-15.75&31.95&0.2219&0.3890&$4.27\cdot 10^{11}$ \\ \hline
SkP&-15.95&30.00&0.2376&0.3812&$4.42\cdot 10^{11}$ \\ \hline
SkRA&-15.78&31.32&0.2631&0.3868&${\bf 6.73\cdot 10^{11}}$ \\ \hline
SkS1&-15.86&28.75&0.2456&0.3772&$4.35\cdot 10^{11}$ \\ \hline
SkS2&-15.89&29.23&0.2424&0.3788&$4.37\cdot 10^{11}$ \\ \hline
SkS3&-15.88&28.84&0.2452&0.3774&$4.36\cdot 10^{11}$ \\ \hline
SkS4&-15.88&28.35&0.2490&0.3755&$4.36\cdot 10^{11}$ \\ \hline
SkSC1&-15.85&28.10&0.2506&0.3747&$4.34\cdot 10^{11}$ \\ \hline
SkSC2&-15.90&24.74&0.2817&0.3591&$4.40\cdot 10^{11}$ \\ \hline
SkSC3&-15.85&27.01&0.2597&0.3715&$4.33\cdot 10^{11}$ \\ \hline
SkSC4&-15.87&28.80&0.2454&0.3773&$4.35\cdot 10^{11}$ \\ \hline
SkSC4o&-15.87&27.00&0.26&0.37&$4.35\cdot 10^{11}$ \\ \hline
SkSC5&-15.85&30.99&0.2294&0.3853&$4.34\cdot 10^{11}$ \\ \hline
SkSC6&-15.92&24.57&0.2837&0.2581&$4.40\cdot 10^{11}$ \\ \hline
SkSC10&-15.96&22.83&0.3034&0.3483&$4.44\cdot 10^{11}$ \\ \hline
SkSC11&-15.87&28.80&0.2454&0.3773&$4.36\cdot 10^{11}$ \\ \hline
SkSC14&-15.92&30.00&0.2372&0.3814&$4.40\cdot 10^{11}$ \\ \hline
SkSC15&-15.88&28.00&0.2518&0.3741&$4.36\cdot 10^{11}$ \\ \hline    
SkSP.1&-15.90&28.00&0.2521&0.3739&$4.38\cdot 10^{11}$ \\ \hline
SkT&-15.40&33.66&0.2073&0.3963&$4.0\cdot 10^{11}$ \\ \hline
SkT1&-15.98&32.02&0.2244&0.3878&$4.46\cdot 10^{11}$ \\ \hline
SkT2&-15.94&32.00&0.2240&0.388&$4.42\cdot 10^{11}$ \\ \hline
SkT3&-15.95&31.50&0.2373&0.3863&$4.43\cdot 10^{11}$ \\ \hline
SkT4&-15.96&35.24&0.2054&0.3973&$4.44\cdot 10^{11}$ \\ \hline
SkT5&-16.00&37.00&0.68&0.4016&$4.48\cdot 10^{11}$ \\ \hline
SkT6&-15.96&29.97&0.2379&0.3811&$4.43\cdot 10^{11}$ \\ \hline
SkT7&-15.94&29.52&0.2409&0.3795&$4.44\cdot 10^{11}$ \\ \hline
SkT8&-15.94&29.92&0.2380&0.381&$4.41\cdot 10^{11}$ \\ \hline
SkT9&-15.88&29.76&0.2384&0.3808&$4.36\cdot 10^{11}$ \\ \hline
\end{tabular}
\newpage
{\begin{tabular}{|l|c|c|c|c|c|}\hline
Skyrme force &$E_{0}$ &$ S_{0}$ &$ \delta$ & x &$\rho_{ND}[\frac{\mbox{g}}{\mbox{cm}^{3}}]$ \\ \hline\hline
$\mbox{SkT1}^{*}$&-16.20&32.31&0.2253&0.3873&$4.64\cdot 10^{11}$ \\ \hline
$\mbox{SkT3}^{*}$&-16.20&31.97&0.275&0.3862&$4.64\cdot 10^{11}$ \\ \hline
SkT1a&-15.98&32.02&0.2244&0.3878&$4.46\cdot 10^{11}$ \\ \hline
SkT2a&-15.94&32.00&0.2240&0.388&$4.42\cdot 10^{11}$ \\ \hline
SkT3a&-15.95&31.50&0.2273&0.38630&$4.42\cdot 10^{11}$ \\ \hline
SkT4a&-15.96&35.45&0.2042&0.3979&$4.44\cdot 10^{11}$ \\ \hline
SkT5a&-16.00&37.00&0.1968&0.4016&$4.48\cdot 10^{11}$ \\ \hline
SkT6a&-15.96&29.97&0.2379&0.3910&$4.43\cdot 10^{11}$ \\ \hline
SkT7a&-15.94&29.52&0.2409&0.3795&$4.41\cdot 10^{11}$ \\ \hline
SkT8a&-15.94&29.92&0.2381&0.381&$4.41\cdot 10^{11}$ \\ \hline
SkT9a&-15.88&29.76&0.2384&0.3808&$4.37\cdot 10^{11}$ \\ \hline
SkTK&-16.70&35.57&0.2122&0.3939&${\bf 5.08\cdot 10^{11}}$ \\ \hline
SKX&-16.05&31.10&0.2313&0.3843&$4.51\cdot 10^{11}$ \\ \hline
SKXce&-15.86&30.15&0.2353&0.3823&$4.35\cdot 10^{11}$ \\ \hline
SKXm&-16.04&31.20&0.2305&0.3847&$4.50\cdot 10^{11}$ \\ \hline
SKxs15&-15.76&31.88&0.2244&0.3878&$4.40\cdot 10^{11}$ \\ \hline
SKxs20&-15.81&35.50&0.2022&0.3989&$4.32\cdot 10^{11}$ \\ \hline
SKxs25&-15.87&39.60&0.1835&0.4082&$4.38\cdot 10^{11}$ \\ \hline
SKz-1&-16.01&32.00&0.2249&0.3875&$4.48\cdot 10^{11}$ \\ \hline
SKz0&-16.01&32.00&0.2249&0.3875&$4.0\cdot 10^{11}$ \\ \hline
SKz1&-16.01&32.01&0.2248&0.3876&$4.48\cdot 10^{11}$ \\ \hline
SKz2&-16.01&32.01&0.2248&0.3876&$4.48\cdot 10^{11}$ \\ \hline
SKz3&-16.01&32.01&0.2248&0.3876&$4.48\cdot 10^{11}$ \\ \hline
SKz4&-16.01&32.01&0.2248&0.3876&$4.48\cdot 10^{11}$ \\ \hline
SLy0&-15.97&31.98&0.2245&0.3877&$4.45\cdot 10^{11}$ \\ \hline
SLy1&-15.99&31.99&0.2247&0.3876&$4.46\cdot 10^{11}$ \\ \hline
SLy2&-15.99&32.00&0.2246&0.3877&$4.46\cdot 10^{11}$ \\ \hline
SLy230a&-15.99&31.99&0.2247&0.3876&$4.46\cdot 10^{11}$ \\ \hline
SLy230b&-15.97&32.01&0.2243&0.3878&$4.44\cdot 10^{11}$ \\ \hline
SLy3&-15.94&31.97&0.2242&0.3879&$4.22\cdot 10^{11}$ \\ \hline
  \end{tabular}
\newpage
\begin{tabular}{|l|c|c|c|c|c|}\hline
Skyrme force &$E_{0}$ &$ S_{0}$ &$ \delta$ & x &$ \rho_{ND}[\frac{\mbox{g}}{\mbox{cm}^{3}}]$ \\ \hline\hline  
SLy4&-15.97&32.00&0.2244&0.3878&$4.45\cdot 10^{11}$ \\ \hline
SLy5&-15.99&32.01&0.2245&0.3877&$4.46\cdot 10^{11}$ \\ \hline
SLy6&-15.92&31.96&0.2240&0.388&$4.40\cdot 10^{11}$ \\ \hline
SLy7&-15.90&31.99&0.2235&0.3882&$4.38\cdot 10^{11}$ \\ \hline
SLy8&-15.97&32.00&0.2244&0.3878&$4.44\cdot 10^{11}$ \\ \hline
SLy9&-15.80&31.98&0.2223&0.3888&$4.30\cdot 10^{11}$ \\ \hline
SLy10&-15.90&31.90&0.2241&0.3879&$4.28\cdot 10^{11}$ \\ \hline
SQMC1&-14.00&29.68&0.2131&0.3934&${\bf 3.00\cdot 10^{11}}$ \\ \hline
SQMC2&-14.29&28.70&0.2239&0.388&${\bf 3.18\cdot 10^{11}}$ \\ \hline
SQMC3&-15.98&{\bf 45.78}&0.1615&0.4192&$4.49\cdot 10^{11}$ \\ \hline
SQMC600&-15.74&34.38&0.2074&0.3963&$4.26\cdot 10^{11}$ \\ \hline
SQMC650&-15.57&33.65&0.2094&0.3953&$4.12\cdot 10^{11}$ \\ \hline
SQMC700&-15.49&33.47&0.2095&0.3952&$4.06\cdot 10^{11}$ \\ \hline
SQMC750&-15.60&33.75&0.2092&0.3954&$4.15\cdot 10^{11}$ \\ \hline
SSK&-16.16&33.50&0.2175&0.3912&$4.61\cdot 10^{11}$ \\ \hline
SV&-16.05&32.82&0.2202&0.3899&$4.51\cdot 10^{11}$ \\ \hline
SV-bas&-15.91&30.00&0.2371&0.3814&$4.39\cdot 10^{11}$ \\ \hline
SV-min&-15.91&30.66&0.2324&0.3838&$4.39\cdot 10^{11}$ \\ \hline
SVI&-15.76&26.88&0.2595&0.3702&$4.27\cdot 10^{11}$ \\ \hline
SVII&-15.79&26.96&0.2592&0.3704&$4.29\cdot 10^{11}$ \\ \hline
SV-K218&-15.90&30.00&0.2369&0.3815&$4.38\cdot 10^{11}$ \\ \hline
SV-K226&-15.90&30.00&0.2369&0.3815&$4.38\cdot 10^{11}$ \\ \hline
SV-K241&-15.91&30.00&0.2371&0.3814&$4.39\cdot 10^{11}$ \\ \hline
SV-kap00&-15.90&30.00&0.2369&0.3815&$4.38\cdot 10^{11}$ \\ \hline
SV-kap02&-15.90&30.00&0.2369&0.3815&$4.38\cdot 10^{11}$ \\ \hline
SV-kap06&-15.91&30.00&0.2371&0.3814&$4.39\cdot 10^{11}$ \\ \hline
SV-mas07&-15.89&30.00&0.2368&0.3816&$4.38\cdot 10^{11}$ \\ \hline
SV-mas08&-15.90&30.00&0.2369&0.3815&$4.38\cdot 10^{11}$ \\ \hline
SV-mas10&-15.91&30.00&0.2371&0.3814&$4.39\cdot 10^{11}$ \\ \hline
SV-sym28&-15.47&28.47&0.2564&0.3718&$4.87\cdot 10^{11}$ \\ \hline
\end{tabular}
\newpage
\begin{tabular}{|l|c|c|c|c|c|}\hline
  Skyrme force &$E_{0}$ &$ S_{0}$ &$ \delta$ & x &$ \rho_{ND}[\frac{\mbox{g}}{\mbox{cm}^{3}}]$ \\ \hline\hline
SV-sym32&-15.94&32.00&0.2240&0.388&$4.42\cdot 10^{11}$ \\ \hline
SV-sym34&-15.97&34.00&0.2123&0.3938&$4.44\cdot 10^{11}$ \\ \hline
SV-tls&-15.89&30.00&0.2367&0.3816&$4.37\cdot 10^{11}$ \\ \hline
T&-15.93&28.35&0.2498&0.3751&$4.41\cdot 10^{11}$ \\ \hline
T11&-16.01&32.00&0.2249&0.3975&$4.44\cdot 10^{11}$ \\ \hline
T12&-16.00&32.00&0.2247&0.3876&$4.68\cdot 10^{11}$ \\ \hline
T13&-16.00&32.00&0.2247&0.3876&$4.68\cdot 10^{11}$ \\ \hline
T14&-15.99&32.00&0.2246&0.3877&$4.46\cdot 10^{11}$ \\ \hline
T15&-16.01&32.00&0.2249&0.3875&$4.48\cdot 10^{11}$ \\ \hline
T16&-16.01&32.00&0.2249&0.3875&$4.48\cdot 10^{11}$ \\ \hline
T21&-16.03&32.00&0.2251&0.3874&$4.49\cdot 10^{11}$ \\ \hline
T22&-16.02&32.00&0.225&0.3875&$4.49\cdot 10^{11}$ \\ \hline
T23&-16.01&32.00&0.2249&0.3875&$4.48\cdot 10^{11}$ \\ \hline
T24&-16.01&32.00&0.2249&0.3875&$4.48\cdot 10^{11}$ \\ \hline
T25&-15.99&32.00&0.2246&0.3877&$4.46\cdot 10^{11}$ \\ \hline
T26&-15.98&32.00&0.2249&0.3875&$4.48\cdot 10^{11}$ \\ \hline
T31&-16.02&32.00&0.225&0.3875&$4.49\cdot 10^{11}$ \\ \hline
T32&-16.03&32.00&0.2251&0.3874&$4.49\cdot 10^{11}$ \\ \hline
T33&-16.02&32.00&0.225&0.3875&$4.49\cdot 10^{11}$ \\ \hline
T34&-16.02&32.00&0.225&0.3875&$4.49\cdot 10^{11}$ \\ \hline
T35&-16.00&32.00&0.2247&0.3876&$4.47\cdot 10^{11}$ \\ \hline
T36&-15.99&32.00&0.2246&0.3877&$4.46\cdot 10^{11}$ \\ \hline
T41&-16.06&32.00&0.2255&0.3872&$4.52\cdot 10^{11}$ \\ \hline
T42&-16.05&32.00&0.2254&0.3873&$4.51\cdot 10^{11}$ \\ \hline
T43&-16.04&32.00&0.2252&0.3874&$4.50\cdot 10^{11}$ \\ \hline
T44&-16.02&32.00&0.225&0.3875&$4.49\cdot 10^{11}$ \\ \hline
T45&-16.02&32.00&0.225&0.3875&$4.49\cdot 10^{11}$ \\ \hline
T46&-16.00&32.00&0.2247&0.3876&$4.47\cdot 10^{11}$ \\ \hline
T51&-16.05&32.00&0.2254&0.3873&$4.46\cdot 10^{11}$ \\ \hline
\end{tabular}
\newpage
\begin{tabular}{|l|c|c|c|c|c|}\hline
Skyrme force &$E_{0}$ &$ S_{0}$ &$ \delta$ & x &$\rho_{ND}[\frac{\mbox{g}}{\mbox{cm}^{3}}]$ \\ \hline\hline
T52&-16.06&32.00&0.2255&0.3872&$4.46\cdot 10^{11}$ \\ \hline
T53&-16.02&32.00&0.225&0.3875&$4.49\cdot 10^{11}$ \\ \hline
T54&-16.03&32.00&0.2251&0.3874&$4.49\cdot 10^{11}$ \\ \hline
T55&-16.03&32.00&0.2251&0.3874&$4.49\cdot 10^{11}$ \\ \hline
T56&-16.01&32.00&0.2249&0.3875&$4.48\cdot 10^{11}$ \\ \hline
T61&-16.07&32.00&0.2256&0.3872&$4.46\cdot 10^{11}$ \\ \hline
T62&-16.07&32.00&0.2256&0.3872&$4.46\cdot 10^{11}$ \\ \hline
T63&-16.06&32.00&0.2255&0.3872&$4.46\cdot 10^{11}$ \\ \hline
T64&-16.03&32.00&0.2251&0.3874&$4.49\cdot 10^{11}$ \\ \hline
T65&-16.04&32.00&0.2252&0.3874&$4.49\cdot 10^{11}$ \\ \hline
T66&-16.02&32.00&0.225&0.3875&$4.49\cdot 10^{11}$ \\ \hline

v070&-15.78&27.98&0.2506&0.3747&${\bf 3.10\cdot 10^{11}}$ \\ \hline
v075&-15.80&28.00&0.2507&0.3746&${\bf 3.11\cdot 10^{11}}$ \\ \hline
v080&-15.79&28.00&0.2506&0.3747&${\bf 3.0\cdot 10^{11}}$ \\ \hline
v090&-15.79&28.00&0.2506&0.3747&${\bf 3.10\cdot 10^{11}}$ \\ \hline
v100&-15.79&28.00&0.2506&0.3747&${\bf 3.10\cdot 10^{11}}$ \\ \hline
v105&-15.79&28.00&0.2506&0.3747&${\bf 3.10\cdot 10^{11}}$ \\ \hline
v110&-15.79&28.00&0.2506&0.3747&${\bf 3.10\cdot 10^{11}}$ \\ \hline
Z&-15.97&26.82&0.2604&0.3698&$4.29\cdot 10^{11}$ \\ \hline
ZR1a&-16.99&9.84&0.65&0.75&${\bf 6.96\cdot 10^{11}}$ \\ \hline
ZR1b&-16.99&18.50&0.385&0.3075&${\bf 5.47\cdot 10^{11}}$ \\ \hline
ZR1c&-16.99&31.50&0.2407&0.3796&${\bf 5.34\cdot 10^{11}}$ \\ \hline
ZR2b&-16.99&11.95&0.5562&0.2219&${\bf 6.16\cdot 10^{11}}$ \\ \hline
ZR2c&-16.99&27.43&0.2725&0.3637&${\bf 5.35\cdot 10^{11}}$ \\ \hline
ZR3a&-16.99&{\bf -138.96}&-0.0631&0.53155&${\bf 5.91\cdot 10^{11}}$ \\ \hline
ZR3b&-16.99&{\bf -100.46}&-0.08847&0.5442&${\bf 6.01\cdot 10^{11}}$ \\ \hline
ZR3c&-16.99&{\bf -42.71}&-0.2239&0.61195&${\bf 6.65\cdot 10^{11}}$ \\ \hline
Zs&-15.88&26.69&0.2629&0.3682&$4.37\cdot 10^{11}$ \\ \hline
$\mbox{Zs}^{*}$&-15.96&28.80&0.2466&0.3767&$4.43\cdot 10^{11}$ \\ \hline
\end{tabular}
\newpage


\section{$E_{tot}/A$ over proton number for densities $(1.0 - 9.0)\cdot 10^{6}\frac{g}{cm^{3}}$ calculated with SkM*}

\begin{figure}[htbp]

 \centering
 \includegraphics[width=0.7\linewidth]{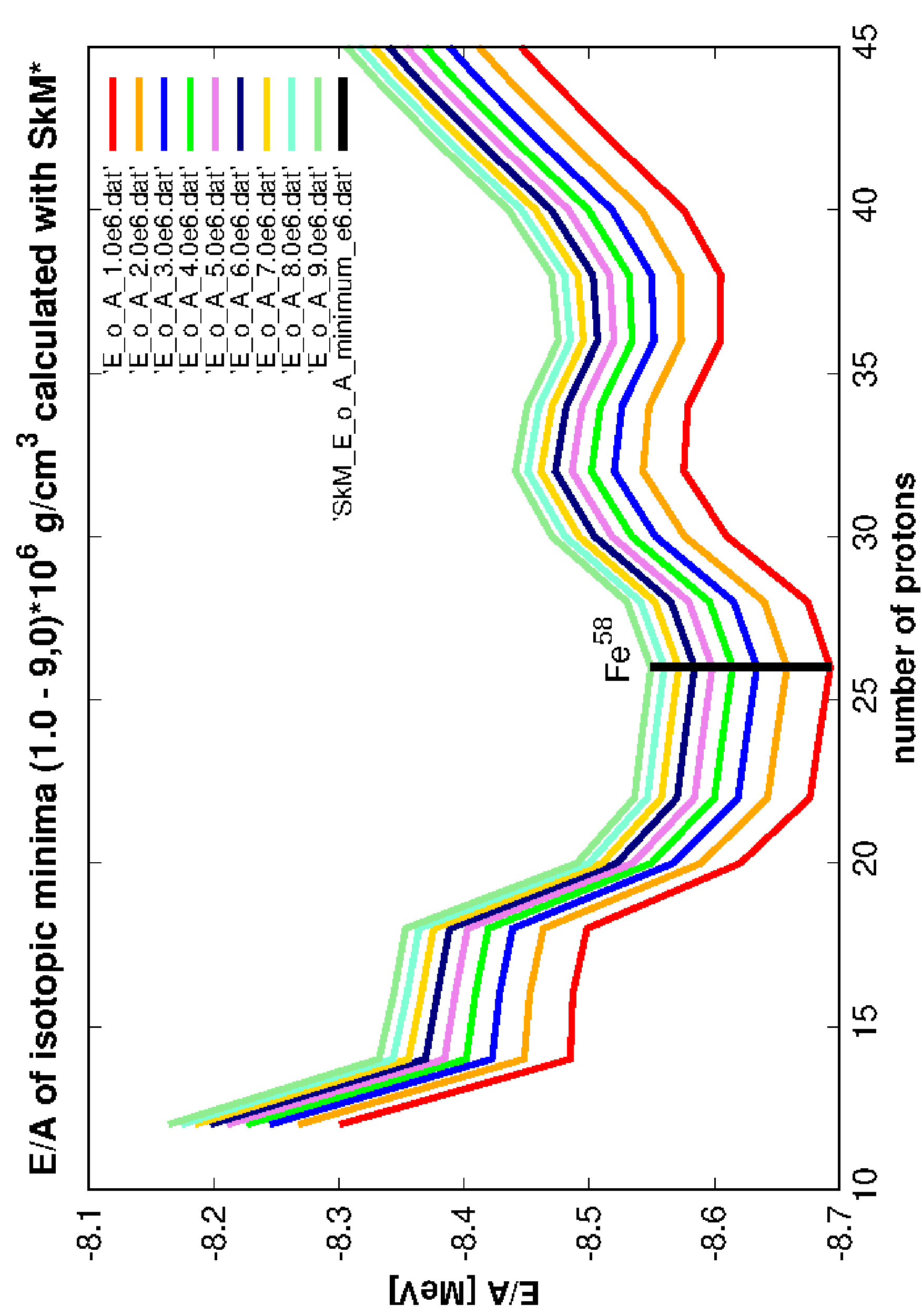}
\caption{\small The binding energies per nucleon E/A of Fe isotopes calculated for baryon densities $(1.0 - 9.0)\cdot 10^{6} \frac{g}{cm^{3}}$ with the SkM* force. As one can see, $\ce Fe^{58}_{26}$ forms the ground state of the outer crust, at these densities.}
\end{figure}
\newpage
\section{$E_{tot}/A$ over proton number for densities $(4.0 - 4.6)\cdot 10^{11}\frac{g}{cm^{3}}$ directly before the drip point (SkM*)}

\begin{figure}[htbp]
\hspace{0.025\linewidth}
 \centering
 \includegraphics[width=0.7\linewidth]{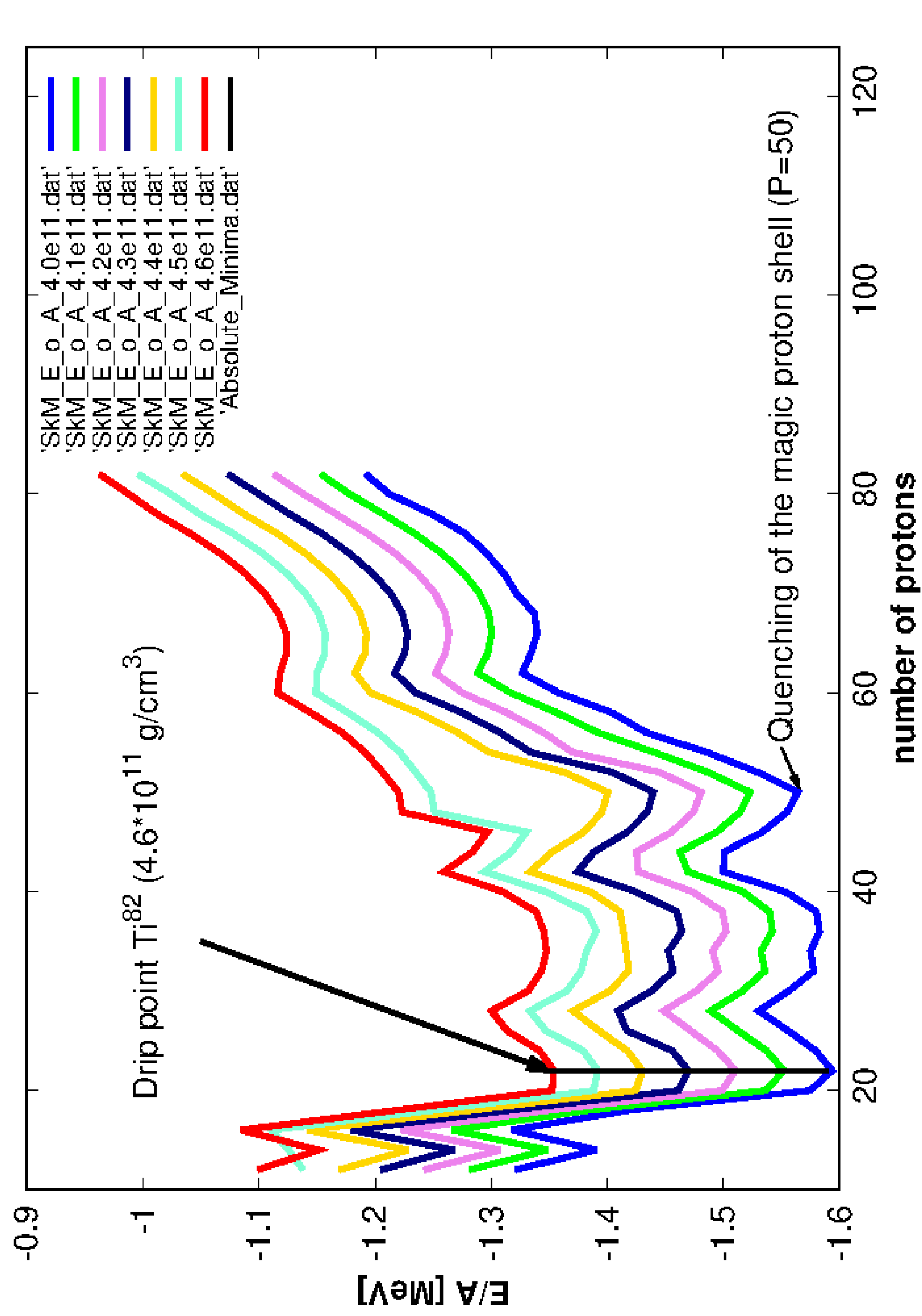}
\caption{\small The binding energy per nucleon as a function of proton number in the drip-region of the outer crust, calculated for the SkM* force. The neutron number of every nucleus shown on the plot minimizes E/A of the isotopic chain.}
\end{figure}
\newpage

\section{Shown are ground state elements over baryon density at the drip point (SkM*)}

\begin{figure}[htbp]
\hspace{0.025\linewidth}
 \centering
 \includegraphics[width=0.7\linewidth]{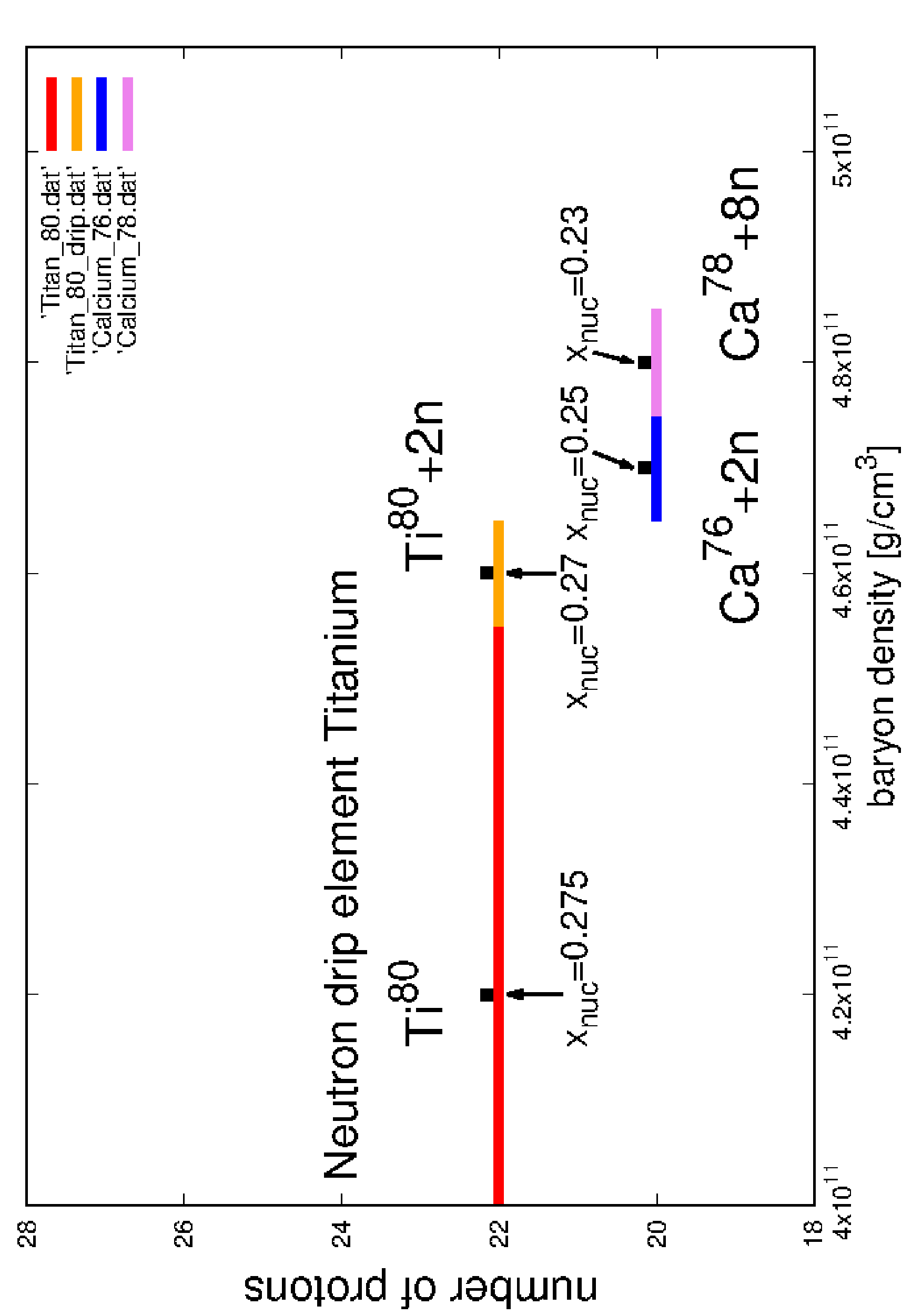}
\caption{\small Neutron drip region calculated with SkM*. Shown are ground states elements over baryon density at the drip point. The drip element $\ce Ti^{82}$ consists of the nuclei $\ce Ti^{80}$ and 2 free neutrons forming a gas. With increasing density the nucleus in the WS-cell gets more and more neutron-rich (decreasing $x_{nuc}$, where only the bound neutrons are taken into account).}
\end{figure}

\end{document}